\documentclass[
10pt,
showpacs,preprintnumbers,nofootinbib,
 amsmath,amssymb,
 aps,
prc,twocolumn,groupedaddress,superscriptaddress,
showkeys
]{revtex4-1}

\usepackage{graphicx}% Include figure files
\usepackage{dcolumn}% Align table columns on decimal point
\usepackage{bm}% bold math
\usepackage[colorlinks=true,urlcolor=blue,citecolor=blue,breaklinks=true]{hyperref}% add hypertext capabilities
\usepackage{epsfig}
\newcommand{\leftsub}[2]{{\vphantom{#2}}_{#1}{#2}}

\newcommand{\bra}[1]{\langle#1|}
\newcommand{\ket}[1]{|#1\rangle}

\newcommand{\cg}[6]{
  C_{#3  #4 \, #5 #6}^{ #1  #2}
}

\begin{document}

%making sub and super scripts smaller:
% \preprint{}

\title{Three-cluster dynamics within the {\em ab initio} no-core shell model with continuum:\\ How many-body correlations and $\alpha$-clustering shape $^6$He}
\author{Sofia Quaglioni}
 \email{quaglioni1@llnl.gov}
 \affiliation{Lawrence Livermore National Laboratory, P.O. Box 808, L-414, Livermore, California 94551, USA}%Lines break automatically or can be forced with \\
\author{Carolina Romero-Redondo}
 \email{romeroredond1@llnl.gov}
 \affiliation{Lawrence Livermore National Laboratory, P.O. Box 808, L-414, Livermore, California 94551, USA}%Lines break automatically or can be forced with \\
 \author{Petr Navr\'atil}
 \email{navratil@triumf.ca}
 \affiliation{TRIUMF, 4004 Wesbrook Mall, Vancouver, British Columbia, V6T 2A3, Canada}%Lines break automatically or can be forced with \\
 \author{Guillaume Hupin}
 \altaffiliation{Present address: Institut de Physique Nucl\'eaire, CNRS/IN2P3, Universit\'e Paris-Sud, Universit\'e Paris-Saclay, F-91406, Orsay, France}
\affiliation{CEA, DAM, DIF, F-91297 Arpajon, France.}
  %\email{---}
 %\affiliation{---}%Lines break automatically or can be forced with \\

\date{\today}% It is always \today, today,

\begin{abstract}
We realize the treatment of bound and continuum nuclear systems in the proximity of a three-body breakup threshold within the {\em ab initio} framework of the no-core shell model with continuum.
Many-body eigenstates obtained from the diagonalization of the Hamiltonian within the harmonic-oscillator expansion of the no-core shell model are coupled with continuous microscopic three-cluster states to correctly describe the nuclear wave function both in the interior
and asymptotic regions. We discuss the formalism in detail and give
algebraic expressions for the case of core+$n$+$n$ systems. %, including 
Using similarity-renormalization-group evolved nucleon-nucleon interactions,  %we study the Borromean $^6$He nucleus. 
we analyze the role of $^4$He+$n$+$n$ clustering and many-body
correlations in the ground and low-lying continuum states of the Borromean
$^6$He nucleus, and study the dependence of the energy spectrum on the
resolution scale of the interaction. We show that %
$^6$He small binding energy and extended radii compatible with experiment can
be obtained simultaneously, without recurring to extrapolations. We
also find that a
significant portion of the ground-state energy and the narrow width of
the first $2^+$ resonance stem from many-body correlations that %, in a
%microscopic cluster picture, 
can be interpreted as core-excitation effects.
\end{abstract}
\maketitle

\section{\label{sec:introduction} Introduction}

Since the first applications to the elastic scattering of nucleons on
$^4$He and
$^{10}$Be~\cite{Nollett2007,Quaglioni2008} roughly
ten years ago, 
large-scale computations combined with new and sophisticated theoretical approaches~\cite{Navratil2016,Elhatisari2015,Hagen2012}
have enabled significant progress in the description of dynamical
processes involving light and medium-mass nuclei within the framework of {\em ab initio} theory, i.e.\ by
solving the many-body quantum-mechanical problem of protons and
neutrons interacting through high-quality nuclear force models.
This resulted in high-fidelity predictions for
nucleon-nucleus~\cite{Hagen2012,Hupin2014,Langhammer2015,Lynn2016,Calci2016}
and deuterium-nucleus~\cite{PhysRevLett.114.212502} clustering phenomena and
scattering properties,
as well as predictive calculations of binary reactions, including the $^3$He$(\alpha,\gamma)^7$Be~\cite{Neff2011,DohetEraly2016}
and $^7$Be$(p,\gamma)^8$B~\cite{Navratil2011379} radiative capture
rates (important for solar astrophysics), and the $^3$H$(d,n)^4$He and $^3$He$(d,p)^4$He
fusion processes~\cite{Navratil2012}. A more recent breakthrough %More
has enabled {\em ab initio} calculations of
$\alpha$-$\alpha$ scattering~\cite{Elhatisari2015}, paving the way %, respectively, to the microscopic
for the description of $\alpha$ scattering and capture reactions
during the helium burning and later evolutionary
phases of massive stars.

One of the main drivers of this progress has been the development of
the no-core shell model with continuum, or
NCSMC~\cite{Baroni2013L,Baroni2013}. This is an {\em ab
initio} framework for the description of the
phenomena of clustering and low-energy nuclear
reactions in light nuclei, which realizes an efficient description of
both the interior and asymptotic configurations of
many-body wave functions.
The approach starts from the wave functions of each of the colliding nuclei and
of the aggregate system, obtained within the {\em ab initio} no-core
shell model (NCSM)~\cite{Navratil2000} by working in a many-body
harmonic oscillator (HO)
basis. It then uses the NCSM static solutions for the aggregate system and continuous `microscopic-cluster'
states, made of pairs of nuclei in relative motion with respect to each other, as an over-complete
basis to describe the full dynamical solution of the system.

Recently, the NCSMC formalism was extended to the treatment of
three-cluster dynamics, laying the groundwork for a comprehensive and unified description of
systems characterized by ternary cluster structures, such as
Borromean halo nuclei~\cite{RevModPhys.76.215}, 
as well as light-nuclei reactions with
three nuclear fragments in either the entrance or exit channels. A few
important examples of such reactions include
the $^4$He$(2n,\gamma)^6$He radiative capture (one of the mechanism
by which stars can overcome the instability of the five- and
eight-nucleon systems and create heavier
nuclei~\cite{doi:10.1146/annurev.nucl.48.1.175}),  and 
the $^3$H$(n,2n)^2$H and $^3$H$(t, 2n)^4$He~\cite{PhysRevLett.109.025003,PhysRevLett.111.052501} fusion rates affecting the neutron spectrum generated in fusion experiments
with deuterium-tritium fuel. 
Within the three-cluster extension of the NCSMC we studied how
many-body correlations and $\alpha$+$n$+$n$ clustering shape the bound
and continuum states of the
Borromean $^6$He nucleus~\cite{rom2016}. 
More limited studies of this same system, based solely on the
three-cluster portion of the NCSMC basis, were previously reported in
Refs.~\cite{Quaglioni2013} and~\cite{Romero-Redondo:2014fya}.   
In this paper, we introduce in detail the general
NCSMC formalism for the description of three-cluster dynamics, 
and present an extended discussion of the results published in Ref.~\cite{rom2016} as well as additional results 
for the $^6$He nucleus.  
 
The paper is organized as follows. In Sec.~\ref{sec:formalism}, we
introduce the NCSMC ansatz for 
systems characterized by
a three-cluster asymptotic behavior, discuss the dynamical
equations, and give the algebraic expressions of the overlap and Hamiltonian couplings
between the discrete and continuous NCSMC basis states for the
particular case of core+$n$+$n$ systems. 
We further discuss the procedure used for the solution of
the three-cluster dynamical equations for bound and scattering states,
and explain how we compute the probability density and matter and
point-proton root-mean-square (rms) radii starting from the obtained 
NCSMC solutions for core+$n$+$n$ systems.  
The application of the approach to describe the ground and
continuum states of the Borromean $^6$He nucleus is presented in
Sec.~\ref{sec:application}. Conclusions are drawn in
Sec.~\ref{sec:conclusions}, and detailed expressions for some of the
most complex derivations are presented in Appendix.

\section{\label{sec:formalism} NCSMC with three-cluster channels}
\subsection{\label{sec:ansatz} Ansatz}
The intrinsic motion in a partial-wave of total angular momentum $J$, parity $\pi$ and isospin $T$ of a system of $A$ nucleons characterized by a three-cluster asymptotic behavior 
\begin{align}
\label{eq:trialwf}
	|\Psi^{J^\pi T}\rangle & =\sum_{\lambda} c^{J^\pi T}_{\lambda}|A\lambda J^{\pi}T\rangle \\ &+ \sum_{\nu} \iint dx \, dy \,   
x^2\, y^2 \, G_{\nu}^{J^\pi T}(x,y) \, %\hat 
{\mathcal A}_\nu\, |\Phi^{J^\pi T}_{\nu x y} \rangle \nonumber \,, 	
	% Here I use G rather than \chi because these are not yet the (orthonormal) Schroedinger wave functions.
\end{align}
where $c^{J^\pi T}_\lambda$ and $G_{\nu}^{J^\pi T}(x,y)$ are discrete and continuous variational amplitudes,  
%of the integration variables $x$ and $y$, 
respectively,  $|A\lambda J^{\pi}T\rangle$ is the $\lambda$-th (antisymmetric) $A$-nucleon eigenstate of the composite system in the $J^\pi T$ channel obtained working within the square-integrable many-body HO basis of the {\em ab initio} NCSM~\cite{Navratil2000},  and
\begin{widetext}
\begin{align}
	 |\Phi^{J^\pi T}_{\nu x y} \rangle  = & 
	\Big[\Big(|A-a_{23}~\alpha_1I_1^{\pi_1}T_1\rangle 
	\left (|a_2\, \alpha_2 I_2^{\pi_2} T_2\rangle |a_3\, \alpha_3 I_3^{\pi_3}T_3\rangle \right)^{(s_{23}T_{23})}\Big)^{(ST)} 
	\left(Y_{\ell_x}(\hat{\eta}_{23})Y_{\ell_y}(\hat{\eta}_{1,23})\right)^{(L)}\Big]^{(J^{\pi}T)} \nonumber \\
	& \times \frac{\delta(x-\eta_{23})}{x\eta_{23}} \frac{\delta(y-\eta_{1,23})}{y\eta_{1,23}}
	\label{eq:3bchannel}	
\end{align}
\end{widetext}
are continuous channel states (first introduced in Ref.~\cite{Quaglioni2013}) describing the organization of the nucleons into three clusters of mass numbers $A-a_{23}$, $a_2$, and $a_3$ ($a_{23}=a_2+a_3 < A$), respectively.
Finally, the operator 
 ${\mathcal A}_\nu$ is an appropriate intercluster antisymmetrizer introduced to guarantee the exact preservation of the Pauli exclusion principle. 
 
\begin{figure}[t]
\includegraphics[width=6cm,clip=,draft=false]{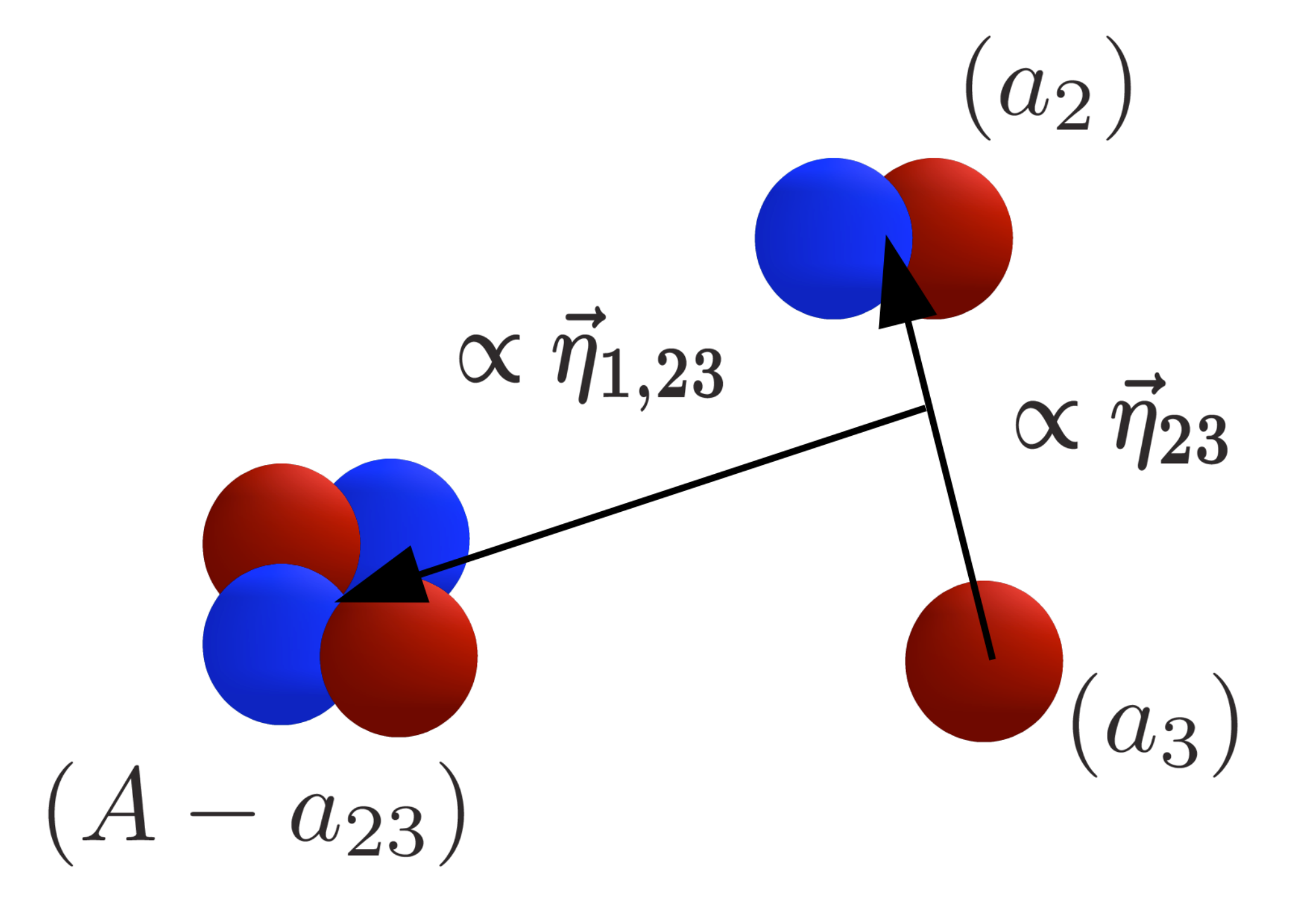}
\caption{(Color online) We show the Jacobi coordinates $\vec\eta_{1,23}$ (proportional to the vector 
between the c.m.\ of the first cluster and that of the
residual two fragments) and $\vec\eta_{23}$ (proportional to the vector between the c.m.\ of 
clusters 2 and 3). In the
figure, a case with three clusters of four, two and one nucleons are shown, however the formalism is completely
general and can be used to describe any three cluster configuration.}
\label{FigCoor}
\end{figure}

In Eq.~\eqref{eq:3bchannel}, $|A-a_{23}~\alpha_1I_1^{\pi_1}T_1\rangle$, $|a_2\, \alpha_2 I_2^{\pi_2} T_2\rangle$ and $|a_3\, \alpha_3 I_3^{\pi_3} T_3\rangle$ represent the  microscopic (antisymmetric) wave functions of the three nuclear fragments, which are also obtained within the NCSM. They are labeled by the angular momentum, parity, isospin and energy quantum numbers $I_i^{\pi_i}$, $T_i$, and $\alpha_i$, respectively, with $i=1,2,3$. Additional quantum numbers characterizing the basis states (\ref{eq:3bchannel}) are the spins $\vec s_{23}=\vec I_2 + \vec I_3$ and $\vec S = \vec I_1+ \vec s_{23}$, the orbital angular momenta $\ell_x$, $\ell_y$ and $\vec L = \vec\ell_x+\vec\ell_y$, and the isospin $\vec T_{23}=\vec T_2+\vec T_3$. In our notation, all these quantum numbers are grouped under the cumulative index $\nu = \{A-a_{23}\, \alpha_1I_1^{\pi_1}T_1;$ $a_2\, \alpha_2 I_2^{\pi_2} T_2;$ $a_3\, \alpha_3 I_3^{\pi_3}T_3;$ $s_{23} \,T_{23}\, S \,\ell_x \,\ell_y \, L\}$. 
Further, the inter-cluster relative motion is described with the help of  the Jacobi coordinates $\vec\eta_{1,23}$ and $\vec\eta_{23}$ where
\begin{align}
	\vec\eta_{1,23} & = \eta_{1,23} \hat{\eta}_{1,23}  \label{eq:etay}\\
	& = \sqrt{\tfrac{a_{23}}{A(A-a_{23})}}  \sum_{i=1}^{A-a_{23}} \vec{r}_i - \sqrt{\tfrac{A-a_{23}}{A\,a_{23}}} \sum_{j=A-a_{23}+1}^A \vec{r}_j \nonumber
\end{align}
is the relative vector proportional to the separation between the center of mass (c.m.) of the first cluster and that of the residual two fragments, and 
\begin{align}
	\vec\eta_{23} & = \eta_{23} \hat{\eta}_{23} \label{eq:etax}\\
	& =\sqrt{\tfrac{a_3}{a_{23}\,a_2}}  \sum_{i=A-a_{23}+1}^{A-a_3} \vec{r}_i - \sqrt{\tfrac{a_2}{a_{23}\,a_3}} \sum_{j=A-a_3+1}^A \vec{r}_j	\nonumber
\end{align}
is the relative coordinate proportional to the distance between the centers of mass of cluster 
2 and 3 (see Fig.~\ref{FigCoor}), where $\vec{r}_i$ denotes the position vector of the $i$-th nucleon.

The NCSM eigenstates appearing in Eqs.~\eqref{eq:trialwf} and \eqref{eq:3bchannel} are obtained by diagonalizing the $A$-, $(A-a_{23})$-, $a_2$-, and $a_3$-nucleon intrinsic Hamiltonians within complete sets of many-body HO basis states, the size of which is defined by the maximum number $N_{\rm max}$ of HO quanta above the lowest configuration shared by the nucleons. The same HO frequency $\hbar\Omega$ is used for the composite nucleus and all three clusters, and the model-space size $N_{\rm max}$ is identical (differs by one) for states of the same (opposite) parity. 
   
The NCSMC ansatz of Eq.~\eqref{eq:trialwf} can be seen as an example of generalized cluster expansion containing single and three-body cluster terms. 
In general such expansion could also contain binary-cluster and/or even higher-body cluster terms, chosen according to  
the particle-decay channels characterizing the system under consideration.  It allows to capture, within a unified consistent framework, both  the single-particle dynamics and microscopic-cluster picture of nuclei. 
For systems in the proximity of a three-body particle-decay channel, but away from two- or higher-body thresholds, Eq.~\eqref{eq:trialwf} 
represents a good ansatz, which converges to the exact solution as $N_{\rm max}\to\infty$. In particular, the square-integrable NCSM eigenstates $|A\lambda J^{\pi}T\rangle$ of the composite nucleus provide an efficient description of the short- to medium-range $A$-body structure of the wave function, while the microscopic three-cluster channels $|\Phi^{J^\pi T}_{\nu x y} \rangle$  make the theory able to handle the long-range and scattering physics of the system. 

\subsection{\label{sec:equations}Dynamical equations}
Adopting the ansatz~\eqref{eq:trialwf} for the many-body wave function and working in the model space spanned by the set of discrete $\ket{A\, \lambda J^\pi T}$ and continuous $%\hat
{\mathcal A}_{\nu} |\Phi^{J^\pi T}_{\nu x y} \rangle$ basis states, the Schr\"odinger equation in each partial wave channel can be mapped onto a generalized eigenvalue problem, schematically given by
\begin{align}
	\left({\bold H} - E\, {\bold N}\right) {\bold C} = 0\,,
%	\left(
%		\begin{array}{c}
%		c \\ \chi
%      	\end{array}
%	\right)  = 0
\label{eq:NCSMC-eq}
\end{align}
where $E$ is the total energy of the system in the c.m.\ reference frame. To simplify the formalism, the specification of the partial wave under consideration ($J^\pi T$ ) is now (and in the remainder of the paper) implied. In Eq.~\eqref{eq:NCSMC-eq} 
\begin{align}
{\bold H}^{\lambda\lambda^\prime}_{\nu x y, \nu^\prime x^\prime y^\prime} = 
	\left(
		\begin{array}{lcl}
      	  E_\lambda\delta_{\lambda\lambda^\prime} & & \bar{h}_{\lambda\nu^\prime} (x^\prime, y^\prime)\\ [2mm] 
		   \bar{h}_{\lambda^\prime\nu}(x,y)  &&\overline{\mathcal{H}}_{\nu\nu^\prime}(x,y,x^\prime, y^\prime) 
		\end{array}
	\right)\,,
	\label{eq:H}
\end{align}
and 
\begin{align}
{\bold N}^{\lambda\lambda^\prime}_{\nu x y, \nu^\prime x^\prime y^\prime} = 
	\left(
		\begin{array}{lcl}
      	  \delta_{\lambda\lambda^\prime} & & \bar{g}_{\lambda\nu^\prime} (x^\prime, y^\prime)\\ [2mm] 
		   \bar{g}_{\lambda^\prime\nu}(x,y)  &&\Delta_{\nu\nu^\prime}(x,y,x^\prime,y^\prime)
		\end{array}
	\right)\,,
	\label{eq:N}
\end{align}
%The
are
two-by-two block-matrices 
representing, respectively, the NCSMC Hamiltonian and norm (or overlap) 
kernels, i.e.\ the matrix elements of the Hamiltonian and identity operators over the set of discrete and continuous basis states spanning the model space. Specifically, the upper diagonal blocks are %built 
NCSM eigenstates of the $A$-nucleon Hamiltonian  
and are trivially given by the diagonal matrix of the  
corresponding eigenenergies $E_\lambda$ and the identity matrix, 
respectively. 
Analogously the lower diagonal blocks 
\begin{align}
\label{eq:formalism_90}
\overline{\mathcal{H}}_{\nu \nu'}(x,y,x^\prime,y^\prime)  &= \left[{\mathcal N}^{-\frac12} %\widetilde{
\mathcal{H}
%}
{\mathcal N}^{-\frac12}\right]_{\nu \nu'}(x,y,x^\prime,y^\prime)\,,\\[2mm]
{\Delta}_{\nu\nu^\prime}(x,y,x^\prime,y^\prime) &= \delta_{\nu\nu^\prime}\frac{\delta(x-x^\prime)}{x x^\prime}\frac{\delta(y-y^\prime)}{y y^\prime}\,,
\end{align} 
are orthonormalized integration kernels obtained from the Hamiltonian  and overlap matrix elements evaluated on the continuous basis states  $%\hat
{\mathcal A}_{\nu} |\Phi^{J^\pi T}_{\nu x y} \rangle$, i.e.\ $\mathcal{H}_{\nu\nu^\prime}(x,y,x^\prime,y^\prime)$ and $\mathcal{N}_{\nu\nu^\prime}(x,y,x^\prime,y^\prime)$. Detailed expressions for these kernels can be found in Ref.~\cite{Quaglioni2013}, 
where we introduced the formalism for the description of three-cluster dynamics based solely on expansions over three-cluster channels states of the type of Eq.~\eqref{eq:3bchannel}. 

The off-diagonal blocks of Eqs.~\eqref{eq:H} and \eqref{eq:N} are given by the couplings between the discrete and continuous sectors of the basis, %appear in the , 
with the cluster form factor, $\bar{g}_{\lambda \nu}(x,y)= [g{\mathcal N}^{-\frac12}]_{\lambda\nu}(x,y)$, and coupling form factor, $\bar{h}_{\lambda \nu}(x,y)= [h{\mathcal N}^{-\frac12}]_{\lambda\nu}(x,y)$, defined in terms of the matrix elements
\begin{align}
g_{\lambda \nu}(x,y) &=  \bra{A\, \lambda J^\pi T} %\hat
{\mathcal A}_{\nu}\ket{\Phi^{J^\pi T}_{\nu x y}}\,,\label{eq:g}\\[2mm]
h_{\lambda \nu}(x,y) &=  \bra{A\, \lambda J^\pi T} %\hat 
H%\hat
{\mathcal A}_{\nu}\ket{\Phi^{J^\pi T}_{\nu x y}}\,, \label{eq:h}
\end{align}
where $%\hat 
H$ is the microscopic $A$-nucleon Hamiltonian.
The general derivation of these three-cluster form factors is outlined in Sec.~\ref{sec:formfactors}, together with their algebraic expressions for the specialized case in which the two lighter fragments are single nucleons.  

Finally, 
\begin{align}
	{\bold C}^\lambda_{\nu x y} = 
		\left(
			\begin{array}{c}
				c_\lambda\\
				\chi_{\nu}(x,y)
			\end{array}	
		\right)
\end{align}
is the vector of the expansion `coefficients', where the relative wave functions $\chi_\nu(x,y)$ are related to the initial unknown continuous amplitudes through 
\begin{equation}
G_\nu(x,y) = [{\mathcal N}^{-\frac12}\chi]_\nu(x,y).   
\label{eq:g_chi}
\end{equation}
These are obtained by solving the NCSMC dynamical equations as discussed in Sec.~\ref{sec:solutions}. %Secs.~\ref{sec:equations} and \ref{sec:solutions}.

\subsection{\label{sec:formfactors} Form factors}
In this  section we discuss in more detail the derivation of the form factors in configuration space introduced in Sec.~\ref{sec:equations}, starting with the coupling form factor $h_{\lambda\nu}(x,y)$ of Eq.~\eqref{eq:h}. 
This can be expressed in terms of the cluster form factor $g_{\lambda\nu}(x,y)$ and three potential form factors
\begin{align}
v^Q_{\lambda \nu}(x,y) & = \bra{A\, \lambda J^\pi T} %\hat
{\mathcal A}_{\nu}%\hat 
{\mathcal V}^Q\ket{\Phi^{J^\pi T}_{\nu x y}}\,,
\label{eq:v}
\end{align}
with $Q$ a generic notation for either $1,23$ or $23$ or $3N$, as
\begin{align}
h_{\lambda \nu}(x,y) & = \left( %\hat 
T_{\rm rel} + \bar V_{\rm C} + E_{\alpha_1} + E_{\alpha_2} + E_{\alpha_3}\right) g_{\lambda \nu}(x,y)\nonumber\\[2mm]
                                 & \phantom{=} +  v^{1,23}_{\lambda\nu}(x,y) + v^{23}_{\lambda\nu}(x,y) + v^{3N}_{\lambda\nu}(x,y)\,. \label{eq:h2}\nonumber\\
\end{align}
The above expression was obtained by separating the microscopic $A$-nucleon Hamiltonian into its relative-motion, average Coulomb and clusters' components according to
\begin{align}
        H = %\hat 
        T_{\rm rel}+\bar V_C+%\hat
        {\mathcal V}_{\rm rel}+%\hat 
        H_{(A-a_{23})}+%\hat 
        H_{(a_2)}+%\hat 
        H_{(a_3)}\,,
        \label{eq:HA}
\end{align}
and taking advantage of the fact that the antisymmetrization operator commutes with $%\hat 
H$. 
$T_{\rm rel}$ is the relative kinetic energy operator for the three-body system, $\bar V_{\rm C} = \bar V^{12}_{\rm C}+ \bar V^{13}_{\rm C}+\bar V^{23}_{\rm C}$ is the sum of the pairwise average Coulomb interactions among the three clusters, and $E_{\alpha_i}$ is the eigenenergy of the $i$-th cluster, obtained by diagonalizing their respective intrinsic Hamiltonians, $H_{(A-a_{23})}$, $H_{(a_2)}$ and $H_{(a_3)}$. Further, $%\hat 
{\mathcal V}_{\rm rel} = %\hat 
{\mathcal V}^{1,23} + %\hat 
{\mathcal V}^{23}+%\hat 
{\mathcal V}^{3N}$ denotes the relative potential, with 
\begin{align}
       {\mathcal V}^{1,23} &= \sum_{i=1}^{A-a_{23}}\sum_{j=A-a_{23}+1}^A\left( V^{NN}_{ij}-\frac{\bar V^{12}_{\rm C}+\bar V^{13}_{\rm C}}{(A-a_{23})a_{23}}\right)\,, \label{eq:interaction1-23}\\[2mm]  
         {\mathcal V}^{23}  &= \sum_{k=A-a_{23}+1}^{A-a_3}\sum_{l=A-a_3+1}^{A} \left(V^{NN}_{kl}-\frac{\bar V^{23}_{\rm C}}{a_2 a_3}\right)\,, \label{eq:interaction23}  
\end{align}
and ${\mathcal V}^{3N}$  the inter-cluster interaction due to the three-nucleon force, which in general is part of a realistic Hamiltonian. In Eqs.~\eqref{eq:interaction1-23} and \eqref{eq:interaction23}, the notation $V^{NN}$ stands for  the nuclear plus point-Coulomb two-body potential.   We note that ${\mathcal V}_{\rm rel}$ is a short-range operator. Indeed, because of the subtraction of $\bar V_{\rm C}$,  the overall Coulomb contribution decreases as the inverse square of the distances between pairs of clusters. 

In the present paper we will consider only the nucleon-nucleon ($NN$) component of the inter-cluster interaction and disregard, for the time being, the term  ${\mathcal V}^{3N}$. The inclusion of the three-nucleon force into the formalism, although computationally much more involved, is straightforward and will be the matter of future investigations. In the remainder of the paper, we will also omit the average Coulomb potential $\bar V_C$, which is null for neutral systems such as the $^4$He+$n$+$n$ investigated here.  The treatment of charged system is nevertheless possible (at least in an approximate way) 
and can be implemented along the same lines of Ref.~\cite{Descouvemont:2005rc}. 

The use of Jacobi coordinates and translational invariant NCSM
eigenstates of the $A$-nucleon system and microscopic-cluster states
represents the `natural' choice for the computation of the
configuration-space form factors of Eqs.~\eqref{eq:g} and
\eqref{eq:v}. However, such a relative-coordinate formalism is only
practical for few-nucleon systems. To access p-shell nuclei, it is
more efficient to work with single-particle coordinates and
Slater-determinant (SD) basis states. As we outline in the following, 
the unique properties of the HO
basis allows us to work with SD functions and still preserve the
translational invariance of the form factors. 
 
In a first step, we compute  matrix elements analogous  to Eqs.~\eqref{eq:g} and \eqref{eq:v} but evaluated in an HO SD basis, i.e.\
\begin{align}
&\leftsub{\rm SD}{\bra{A\, \lambda J^\pi T} %\hat
{\mathcal O}_{\rm t.i.}\ket{\Phi^{J^\pi T}_{\gamma n_x n_y}}}_{\rm SD}\,,\label{eq:SDme} 
\end{align}
where $%\hat
{\mathcal O}_{\rm t.i.} = %\hat
{\mathcal A}_\nu,\, %\hat
{\mathcal A}_\nu %\hat 
{\mathcal V}^{1,23},\, %\hat
{\mathcal A}_\nu %\hat 
{\mathcal V}^{23}$ is a translational invariant operator. The SD NCSM
eigenstates of the $A$-nucleon system factorize into the product of their translational-invariant counterparts with the $0\hbar\Omega$ HO motion of their c.m.\  coordinate $\vec{R}^{(A)}_{\rm c.m.}$,
\begin{align}
	\ket{A \lambda J^\pi T}_{\rm SD} = \ket{A \lambda J^\pi T} \,R_{00}({R}^{(A)}_{\rm c.m.})Y_{00}(\hat{R}^{(A)}_{\rm c.m.})\,. \label{SD-eigenstate}
\end{align}
At the same time, the kets in Eq.~\eqref{eq:SDme} are a set of HO three-cluster channel states, defined as 
 \begin{widetext}
\begin{align}
|\Phi_{\gamma n_x n_y}^{J^{\pi}T}\rangle_{\rm SD}= &\left[\left(|A-a_{23}~\alpha_1 I_1^{\pi_1}T_1\rangle_{\rm SD} \left(Y_{\ell_x}(\hat{\eta}_{23}) 
\left(|a_2\alpha_2I_2^{\pi_2}T_2\rangle |a_3\alpha_3I_3^{\pi_3}T_3\rangle\right)^{( s_{23} T_{23} )}\right)^{(J_{23}T_{23})}\right)^{(ZT)} 
Y_{\ell_y}(\hat{R}^{a_{23}}_{\rm c.m.})\right]^{(J^{\pi}T)} \nonumber\\[2mm]
&\times R_{n_x\ell_x}(\eta_{23})R_{n_y\ell_y}({R}^{a_{23}}_{\rm c.m.})\,,
\label{eq:3bchannelSD}
\end{align}
\end{widetext}
describing the motion of the heaviest of the two clusters and of the system formed by the second and third clusters in the `laboratory' reference frame. Here  
\begin{align}
	&\vec R^{(A-a_{23})}_{\rm c.m.} = R^{(A-a_{23})}_{\rm c.m.} \hat R^{(A-a_{23})}_{\rm c.m.}= \frac{1}{\sqrt{A-a_{23}}}\sum_{i=1}^{A-a_{23}}\vec r_i\,,\\
      &\vec R^{(a_{23})}_{\rm c.m.}    =R^{(a_{23})}_{\rm c.m.} \hat R^{(a_{23})}_{\rm c.m.}= \frac{1}{\sqrt{a_{23}}}\sum_{j=A-a_{23}+1}^A \vec r_j\,.
\end{align}
are respectively the coordinates of the c.m.\ of the first and last two clusters,  
$|A-a_{23}~\alpha_1 I_1^{\pi_1}T_1\rangle_{\rm SD}$ are the SD NCSM eigenstates of the $(A-a_{23})$-nucleon system, i.e.\
\begin{align}
&|A-a_{23}~\alpha_1 I_1^{\pi_1}T_1\rangle_{\rm SD}\\
&\quad =|A-a_{23}~\alpha_1 I_1^{\pi_1}T_1\rangle\, R_{00}(R^{(A-a_{23})}_{\rm c.m.})Y_{00}(\hat R^{(A-a_{23})}_{\rm c.m.})\,,\nonumber
\end{align}
and $R_{n_x\ell_x}(\eta_{23})$ and $R_{n_y\ell_y}({R}^{a_{23}}_{\rm
  c.m.})$ are HO radial wave functions. 

The HO channel states of Eq.~\eqref{eq:3bchannelSD} differ from the
original basis of Eq.~\eqref{eq:3bchannel} also in the angular
momentum coupling scheme, as reflected in the new channel index
$\gamma = \{A-a_{23}\, \alpha_1I_1^{\pi_1}T_1; $ $a_2\, \alpha_2
I_2^{\pi_2} T_2; $$a_3\, \alpha_3 I_3^{\pi_3}T_3; \ell_x
\,s_{23}J_{23} \,T_{23}\, Z \,\ell_y\}$.  
Here $J_{23}$ denotes the total (orbital plus spin) angular momentum quantum number of the system formed by the second and third clusters and $\vec Z = \vec I_1 + \vec J_{23}$ a channel spin. The use of different coupling schemes is purely dictated by convenience, as it will become apparent from Secs.~\ref{sec:Am211} and \ref{sec:solutions} where we discuss, respectively, the derivation of the matrix elements~\eqref{eq:SDme} in the special instance of a core nucleus plus two single nucleons ($a_2,a_3=1$), and the solution of the NCSMC dynamical equations.

Both the states of Eqs.~\eqref{SD-eigenstate} and
\eqref{eq:3bchannelSD} contain the spurious motion of the center of
mass. However, by exploiting the orthogonal transformation between the pairs of coordinates $\{\vec R^{(A-a_{23})}_{\rm c.m.},\vec R^{(a_{23})}_{\rm c.m.}\}$ and $\{\vec R^{(A)}_{\rm c.m.},\vec\eta_{1,23}\}$, and performing the transformation to the angular momentum coupling scheme of Eq.~\eqref{eq:3bchannel}  we recover the purely translationally-invariant matrix elements over the original channel states~\eqref{eq:3bchannel}, i.e.\
\begin{align}
        \bra{A \lambda J^\pi T} %\hat
        {\mathcal O}_{\rm t.i.}\ket{\Phi^{J^\pi T}_{\nu x y}}  = 
        &\!\sum_{n_x n_y} \!\!R_{n_x\ell_x}(x)R_{n_y\ell_y}(y)\!\!\sum_{Z J_{23}} \hat Z \hat J_{23}\hat S \hat L \nonumber\\
        &\!\times (-1)^{I_1+J_{23}+J+S+Z+\ell_x+\ell_y}\nonumber \\
        &\!\times 
        \left\{
                \begin{array}{ccc}
                        I_1 & s_{23} & S\\[2mm]
                        \ell_x & Z & J_{23}\\[2mm]
                \end{array}
        \right\}
        \left\{
                \begin{array}{ccc}
                        S & \ell_x & Z\\[2mm]
                        \ell_y & J & L\\[2mm]
                \end{array}
        \right\}\nonumber\\[2mm]
        & \!\times\frac{\leftsub{\rm SD}{\bra{A\, \lambda J^\pi T} %\hat
        {\mathcal O}_{\rm t.i.}\ket{\Phi^{J^\pi T}_{\gamma n_x n_y}}}_{\rm SD}}{\bra{n_y\, \ell_y\, 0\, 0\, \ell_y\,}\,0\,0\,n_y\,\ell_y\,\ell_y\rangle_{\frac{a_{23}}{A-a_{23}}}}. \nonumber\\[2mm]
        &\label{eq:transf}
\end{align}
Here,
$\hat Z = \sqrt{2Z+1}, \cdots$ etc., the generalized HO bracket due to the c.m.\ motion is simply given by
\begin{align}
	\bra{n_y\, \ell_y\, 0\, 0\, \ell_y\,}\,0\,0\,n_y\,\ell_y\,\ell_y\rangle_{\frac{a_{23}}{A-a_{23}}}\! =\!(-1)^{\ell_y}\!\!\left(\frac{a_{23}}{A-a_{23}}\right)\!\!^{\frac{2n_y+\ell_y}{2}},
\end{align}
and we made use of the closure properties of the HO radial wave functions to represent the Dirac's $\delta$-function of Eq.~\eqref{eq:3bchannel}.
Indeed, due to the finite range of the square-integrable $A$-nucleon basis states  $\ket{A\, \lambda J^\pi T}$, the configuration-space matrix elements of the translational invariant operators $%\hat
{\mathcal A}_\nu$ and  $%\hat 
H%\hat
{\mathcal A}_\nu$ of Eqs.~\eqref{eq:g} and \eqref{eq:h} are localized and can be evaluated within an HO model space.

\subsubsection{Matrix elements for $core$+$n$+$n$ systems} 
\label{sec:Am211}
In this section we give an example of how SD form-factor matrix
elements of the type of Eq.~\eqref{eq:SDme} can be derived working
within the second quantization formalism. We do this for the special
case in which, both in the initial and in the final state, two out of
the three clusters are single neutrons (such as the $^4$He+$n$+$n$
system investigated in this paper), and in particular we choose $a_2,
a_3 = 1$. 

As pointed out in Sec.\ II.E.1 of Ref.~\cite{Quaglioni2013}, in such a
case it is convenient to incorporate the trivial antisymmetrization
for the exchange of nucleons $A-1$ and $A$ in the definition of the
channel basis of Eq.~\eqref{eq:3bchannel}. This is simply accomplished
by selecting only the states for which $(-1)^{\ell_x+s_{23}+T_{23}} =
-1$. The inter-cluster antisymmetrizer then reduces to the anstisymmetrization operator for a binary $(A-2,2)$ mass partition, $%\hat
{\mathcal A}_{(A-2,2)}$ (see, e.g.,\ Eq.\ (4) of Ref.~\cite{Navratil2011}). 

Further, it is useful to introduce a channel basis defined entirely in
single-particle coordinates, i.e. 
\begin{align}
        |\Phi^{J^\pi T}_{\kappa_{ab}}\rangle_{\rm SD} = & \Big [\left|A-2\, \alpha_1 I_1 T_1\right\rangle_{\rm SD} \nonumber\\
        &  \times
        \left(\ket{n_a \ell_a j_a \tfrac12} \ket{n_b \ell_b j_b \tfrac12}\right)^{(I T_{23})}\Big]^{(J^\pi T)}\,.
        \label{eq:SD-basis-ab}
\end{align}
Here, $\ket{n_a \ell_a j_a \tfrac12}$ and $\ket{n_b \ell_b j_b \tfrac12}$ are single-particle HO states of nucleon $A$ and $A-1$, respectively, and $\kappa_{ab}=\{A-2 \alpha_1 I^{\pi_1}_1 T_1; $ $n_a\ell_aj_a\tfrac12;n_b\ell_bj_b\tfrac12;IT_{23}\}$. Within this basis, the matrix elements of the translational-invariant operators $%\hat
{\mathcal O}_{\rm t.i.} = %\hat
{\mathcal A}_{(A-2,2)}$, and $%\hat
{\mathcal A}_{(A-2,2)} %\hat
{\mathcal V}^{1,23}$ can be easily obtained in the second quantization
formalism, and the corresponding SD matrix elements of
Eq.~\eqref{eq:SDme} can then be recovered by means of a linear
transformation as described in detail in Sec.\ II.E.1 of
Ref.~\cite{Quaglioni2013}. 

Taking into account that the application of $%\hat
{\mathcal A}_{(A-2,2)}$ on the fully antisymmetric $A$-nucleon bra
simply yields the square root of the binomial coefficient
$\tiny\left(\begin{array}{c} A\\ 2\end{array}\right)$, we then obtain 
\begin{align}
        &\leftsub{\rm SD}{\left\langle A\,\lambda\, J^{\pi} T\left| %\hat
        {\mathcal A}^2_{(A-2,2)}\right | \Phi^{J^\pi T}_{\kappa_{ab}} \right\rangle}_{\rm SD}\nonumber\\[2mm]
        &\qquad= \frac{1}{\sqrt 2} \sum_{\substack{M_{I_1}M_{I}\\m_{j_a}m_{j_b}}} \cg{J}{M}{I_1}{M_{I_1}}{I}{M_{I}} \nonumber\\[2mm]
        &\qquad \times \cg{I}{M_{I}}{j_a}{m_{j_a}}{j_b}{m_{j_b}}\cg{T}{M_T}{T_1}{M_{T_1}}{T_{23}}{M_{T_{23}}}\cg{T_{23}}{M_{T_{23}}}{\tfrac12}{m_{t_a}}{\tfrac12}{m_{t_b}} \nonumber\\[2mm]
        &\qquad \times  \leftsub{\rm SD}{\langle} A \lambda J^\pi T | a^\dag_{i_{ a}} a^\dag_{i_{b}} | A{-}2 \alpha_1 I_1^{\pi_1}T_1\rangle_{\rm SD}\,,
        \end{align}
and
\begin{align}
        &\leftsub{\rm SD}{\left\langle A\,\lambda\, J^{\pi} T \left| %\hat
        {\mathcal A}_{(A-2,2)} %\hat
        {\mathcal V}^{1,23}\right |  \Phi^{J^\pi T}_{\kappa_{ab}}   \right\rangle}_{\rm SD}\nonumber\\[2mm]
        &\qquad=-\frac{1}{\sqrt 2} \sum_{\substack{i_{\bar a}i_{\bar b}i_{\bar c}i_{\bar c'}\\M_{I_1}M_{I}\\m_{j_a}m_{j_b}}} \langle  i_{\bar a} i_{\bar c}  | V^{NN} |  i_{\bar c'} i_{a} \rangle \cg{J}{M}{I_1}{M_{I_1}}{I}{M_{I}} \nonumber\\[2mm] 
        &\qquad \times \cg{I}{M_{I}}{j_a}{m_{j_a}}{j_b}{m_{j_b}}\cg{T}{M_T}{T_1}{M_{T_1}}{T_{23}}{M_{T_{23}}}\cg{T_{23}}{M_{T_{23}}}{\tfrac12}{m_{t_a}}{\tfrac12}{m_{t_b}} \nonumber\\[2mm]
        &\qquad \times \leftsub{\rm SD}{\langle} A \lambda J^\pi T | a^\dag_{i_{\bar a}} a^\dag_{i_{\bar c}} a^\dag_{i_{b}} a_{i_{\bar c'}}| A{-}2 \alpha_1 I_1^{\pi_1}T_1\rangle_{\rm SD}\,,
\label{eq:AV123}
\end{align}
where $\cg{J}{M}{j_1}{m_{j_1}}{j_2}{m_{j_2}}$ are Clebsch-Gordan coefficients, 
$a^\dag$ and $a$ are creation and annihilation operators,
respectively, and $i_{q} = \{n_q \ell_q j_q m_{j_q}\tfrac12 m_{t_q}\}$ are
single-particle quantum numbers. Note that in Eq.~\eqref{eq:AV123}, 
there are summations over the indexes
$i_{\bar q}$ and the bar is only meant to differentiate them better 
from the the ones that correspond to the matrix element being calculated, i.e.,
from $i_a$ and $i_b$.

The above matrix elements 
had already being derived and utilized in the computation of the cluster and coupling form factors required for the unified description of $^6$Li structure and $d$+$^4$He dynamics with chiral two- and three-nucleon forces~\cite{PhysRevLett.114.212502}, as well as in the description of $d+^7$Li scattering based on a high-precision $NN$ potential~\cite{Raimondi2016}. Here we present for the first time their algebraic expressions.

Finally, different from the NCSMC formalism for the description of deuterium-nucleus collisions, where the dynamics of the last two nucleons is already taken into account in the calculation of the (bound) deuterium eigenstates, to obtain the three-cluster coupling form factor of Eq.\eqref{eq:h2}  one has also to compute the potential form factor $v^{23}_{\lambda\nu}(x,y)$ due to the $%\hat
{\mathcal V}^{23}$ interaction of Eq.~\eqref{eq:interaction23}. In the present (neutral) example this is simply given by the action of the operator $V_{A-1,A}=V(x)$ on the cluster form factor, i.e.,
\begin{align}
	v^{23}_{\lambda\nu}(x,y) &= V(x)\, g_{\lambda\nu}(x,y)\,.
\end{align}

\subsection{\label{sec:solutions} Solution of the dynamical equations}
Rather than solving directly Eq.~\eqref{eq:NCSMC-eq} we prefer to work with the set of Schr\"odinger equations
\begin{equation}\label{eq:orthoEq}
\left(  \overline{\bold H} -E \right) 
\overline{\bold C} = 0\,,
\end{equation}
where $\overline{\bold H}$ is the orthogonalized NCSMC Hamiltonian,
\begin{align}\label{eq:orthoH}
  \overline{\bold H}^{\lambda\lambda^\prime}_{\nu xy,\nu^\prime x^\prime y^\prime} &=
  \left[{\bold N}^{-\frac{1}{2}} \,
{\bold H}\,
  {\bold N}^{-\frac{1}{2}}\right]^{\lambda\lambda^\prime}_{\nu xy,\nu^\prime x^\prime y^\prime} \nonumber\\[2mm] 
 & =   \left(
     		\begin{array}{lcl}
			\overline{\bold H}^{(11)}_{\lambda\lambda^\prime} && \overline{\bold H}^{(12)}_{\lambda\nu^\prime}( x^\prime, y^\prime) \\ [2mm]
			\overline{\bold H}^{(21)}_{\lambda^\prime\nu}(x,y) && \overline{\bold H}^{(22)}_{\nu  \nu^\prime}(x, y, x^\prime, y^\prime) \\ [2mm]
		\end{array}
  \right)\,,
\end{align}
${\bold N}^{-\frac 12}$ is the inverse square root of the norm kernel of Eq.~\eqref{eq:N}, and the orthonormal wave functions are given by 
\begin{align}
	\label{eq:orthowf}
	\overline{\bold C}^{\lambda^\prime}_{\nu^\prime x^\prime y^\prime} & = \left[{\bold N}^{\frac12} {\bold C}\right]^{\lambda^\prime}_{\nu^\prime x^\prime y^\prime}%\\[2mm]
%	                            & =
						  = \left(
     							\begin{array}{c}
          							\overline{c}_{\lambda^\prime} \\
		  						\overline\chi_{\nu^\prime}(x^\prime,y^\prime)
     							\end{array}
 						 \right)\,.
\end{align}	
Detailed expressions of ${\bold N}^{-\frac12}$ and of the elements of the orthogonalized Hamiltonian kernel and wave function of of Eqs.~\eqref{eq:orthoH} and \eqref{eq:orthowf}, respectively,  can be found in Appendix~\ref{ap:kernels}.

Further, we introduce the set of hyperspherical coordinates 
\begin{align}
\rho = \sqrt{x^2+y^2}\,\,\quad \mbox{and}\,\,\quad \alpha = \arctan{\frac xy}, 
\label{eq:coordinates} 
\end{align}
and reformulate 
Eq.~\eqref{eq:orthoEq} by taking advantage of the closure and orthogonality properties of the complete set of functions (see also Appendix B and Sec.\ II.C of Ref.~\cite{Quaglioni2013})
\begin{equation}
\phi_K^{\ell_x,\ell_y}(\alpha)=N_K^{\ell_x\ell_y}(\sin\alpha)^{\ell_x} (\cos \alpha)^{\ell_y}
P_n^{\ell_x+\frac{1}{2},\ell_y+\frac{1}{2}}(\cos 2\alpha)\,.
\label{eq:phi}
\end{equation}
Together with the bipolar spherical harmonics $(Y_{\ell_x}(\hat x)
Y_{\ell_y}(\hat y))^{(L)}_{M_L}$, these form the hyperspherical harmonics functions 
\begin{equation}
\mathcal{Y}^{K \ell_x\ell_y}_{L M_L}(\alpha,\hat x,\hat y)=\phi_K^{\ell_x,\ell_y}(\alpha)
\left(Y_{\ell_x}(\hat x)\, Y_{\ell_y}(\hat y)\right)^{(L)}_{M_L}\,,
\label{HHbasis}
\end{equation}
i.e.,\ the eigenfunctions with eigenvalue $K(K+4)$ of the grand-angular part of the relative kinetic energy operator for a three-body system. 
In the definition of Eq.~\eqref{eq:phi}, $P_n^{\alpha,\beta}(\xi)$ are Jacobi polynomials, $N_K^{\ell_x\ell_y}$ normalization constants, and $K = 2n+\ell_x+\ell_y$, with $n$ a positive integer, is the hypermomentum quantum number.  Specifically, by $i)$~using the expansion  
\begin{align}
\overline{\chi}_{\nu^\prime}(\rho^\prime,\alpha^\prime)=\frac{1}{\rho^{\prime 5/2}}\sum_{K^\prime} u_{\nu^\prime K^\prime}(\rho^\prime)
\phi_{K^\prime}^{\ell^\prime_x,\ell^\prime_y}(\alpha^\prime)\,
\label{eq:expansion}
\end{align}
for the orthogonalized continuous amplitudes, $ii)$~multiplying the lower block of Eq.~\eqref{eq:orthoEq} by $\phi_K^{\ell_x,\ell_y}(\alpha)$, and $iii)$~performing all integrations over the hyperangular variables $\alpha$ and $\alpha^\prime$, we arrive at the
set of coupled Bloch-Schr\"odinger equations
\begin{widetext}
\begin{align}
	\left\{
	\begin{array}{l}
		\sum\limits_{\lambda^\prime}\overline{\bold H}^{(11)}_{\lambda\lambda^\prime} \, \overline{c}_{\lambda^\prime}
	     + \sum\limits_{\nu^\prime K^\prime} \int d\rho^\prime \rho^{\prime 5/2}\, \overline{\bold H}^{(12)}_{\lambda\nu^\prime \!K^\prime}(\rho^\prime)\, u_{\nu^\prime \!K^\prime}(\rho^\prime) - E\, \overline{c}_{\lambda}= 0\\[4mm]
		\sum\limits_{\lambda^\prime}\overline{\bold H}^{(21)}_{\lambda^\prime\nu K}(\rho)\, \overline{c}_{\lambda^\prime}	
		+ \sum\limits_{\nu^\prime K^\prime}\int d\rho^\prime \rho^{\prime 5/2} \, \overline{\bold H}^{(22)}_{\nu K, \nu^\prime\! K^\prime}(\rho,\rho^\prime) \,u_{\nu^\prime \!K^\prime}(\rho^\prime) + \left({\mathcal L}_{\nu K}(\rho) - E\right)\, \rho^{ -5/2}\, u_{\nu \!K}(\rho) = {\mathcal L}_{\nu K}(\rho)\, \rho^{ -5/2}\, u_{\nu \!K}^{\rm ext}(\rho)\,.
	\end{array}
	\right .
	\label{eq:system}	
\end{align}
\end{widetext}
Here, the elements of the orthogonalized Hamiltonian kernel in the in the hyperradial variables are given by
\begin{align}
	\overline{\bold H}^{(12)}_{\lambda\nu K}(\rho) & = \overline{\bold H}^{\dagger (21)}_{\lambda\nu K}(\rho) \\
	& =\! \int \!d\alpha (\sin\alpha)^2(\cos\alpha)^2 \phi_K^{*\ell_x,\ell_y}(\alpha)  \overline{\bold H}^{(12)}_{\lambda\nu}(\rho,\alpha)\nonumber
\end{align}
and
\begin{align}
		\overline{\bold H}^{(22)}_{\nu K,\nu^\prime\! K^\prime}(\rho,\rho^\prime) \!=\! & \!\!\iint \!\!d\alpha\, d\alpha^\prime\!(\sin\alpha)^2(\cos\alpha)^2   (\sin\alpha^\prime)^2(\cos\alpha^\prime)^2 \nonumber\\
	& \times \phi_K^{*\ell_x,\ell_y}(\alpha) \overline{\bold H}^{(22)}_{\nu\nu^\prime}(\rho,\alpha,\rho^\prime,\alpha^\prime) \phi_K^{\prime\ell^\prime_x,\ell^\prime_y}(\alpha^\prime)\,.\nonumber\\
\end{align}
To arrive at Eq.~\eqref{eq:system}  we have also divided the configuration space into two regions by assuming that the Coulomb interaction (if present) is the only interaction experienced by the clusters beyond the hyperradius $\rho=a$ (i.e., in the external region), 
and re-framed  the three-cluster problem within the microscopic $R$-matrix formalism~\cite{R-matrix}. 
This is accomplished by adding to and subtracting from the Hamiltonian matrix the operator ${\bold L}$ defined by the two-by-two block matrix 
\begin{align}
	{\bold L}^\lambda_{\nu K\rho} =  
	\left(
		\begin{array}{lcl}
			0 & & 0\\[2mm]
			0 & & {\mathcal L}_{\nu K}(\rho)
		\end{array}	
	\right)\,,
	\label{eq:blochgen}
\end{align}
 where the lower-diagonal block is given by the %usual 
Bloch surface operator ($L_{\nu K}$ being arbitrary constants), 
\begin{align}
\mathcal{L}_{\nu K}(\rho)=\frac{\hbar^2}{2m}\delta(\rho-a)\frac{1}{\rho^{5/2}}\left(\frac{\partial}{\partial\rho}-\frac{L_{\nu K}}{\rho}\right)\rho^{5/2}\,.
\label{eq:bloch}
\end{align}
The operator of Eq.~\eqref{eq:bloch} allows one to conveniently implement the matching between
internal and external solutions at the hyperradius $\rho=a$, and has
the further functions of restoring the hermiticity of the Hamiltonian
matrix in the internal region and enforcing the continuity of the the
derivative of the hyperradial wave function at the matching hyperradius. 
Provided that the matching hyperradius $a$ lies outside of the short-to-mid range
where the discrete $\ket{A\lambda J^\pi T}$ basis states contribute, only the continuous component of the NCSMC wave function is present in the external region. 
Therefore, to find the solutions of the three-cluster NCSMC equations it is sufficient to match the hyperradial wave function $u_{\nu K}(\rho)$ entering Eq.~\eqref{eq:expansion} with the known exact solutions of the three-body Schr\"odinger equation in the external region. For bound states of three-body neutral systems (such as the one investigated in this paper) these are entirely described by the hyperradial wave functions
\begin{align}
	u^{\rm ext}_{\nu K}(\rho) = B_{\nu K}\,\sqrt{k\rho}\,K_{K+2}(k\rho)\,,
	\label{eq:ext}
\end{align}
where $K_{K+2}(k\rho)$ are modified Bessel functions of the second kind, $k^2=-2mE/\hbar^2$ is the wave number, and $B_{K\nu}$ are constants. 
The study of continuum states requires the use of a different set
of external wave functions 
 \begin{align}
u^{J^{\pi}T}_{K\nu}(\rho) \propto\left[ 
H^-_{K}(k\rho)\delta_{\nu\nu'}\delta_{KK'}
-S_{\nu K,\nu' K'}H^+_{K}(k\rho)\right]
	\label{eq:ext_cont}
\end{align}
with $H^{\pm}$ being the incoming
and outgoing functions for neutral systems \cite{Descouvemont:2005rc}, 
%$\kappa$ the wave number,  
and $S$ the three-body scattering matrix of 
the process.

Finally, the discrete coefficients $\overline{c}_\lambda$ and hyperradial wave functions $u_{\nu K}(\rho)$ can be conveniently obtained by applying to Eq.~\eqref{eq:system} the Lagrange-mesh method~\cite{PhysRevA.65.052710,BayeJPB98,Hesse2002184,Hesse199837,Descouvemont:2003ys}, in an analogous way to that presented in Sec.\ II.D and Appendix C of Ref.~\cite{Quaglioni2013}. 

\subsection{Probability density}
\label{sec:prob_dist}

For a three-body system it is useful to define the probability density 
in terms of the Jacobi coordinates of Eqs.~\eqref{eq:etay} and \eqref{eq:etax}. 
This provides 
 a convenient visual description of the 
 distribution of the clusters with respect to one another.
In particular, it highlights
which configuration or configurations are preferred by the system.

In general, this probability density is given by
\begin{equation}
P(x,y)=x^2y^2 |\langle \Psi^{J^\pi T}|\delta(x-\eta_{23})\delta(y-\eta_{1,23})|\Psi^{J^\pi T}  
\rangle|^2.
\label{eq:prob}
\end{equation} 

However, given that the NCSMC wave function contains not only a cluster 
part but also a many-body contribution,   
in our formalism the probability density of Eq.~\eqref{eq:prob} is computed
in an 
approximate way. We project the whole
wave function into the cluster basis, i.e.,  

\begin{widetext} 
\begin{align} 
&|\Psi_{3B}^{J^\pi T}\rangle
 = \sum_{\nu} \iint dx \, dy \,   
x^2\, y^2 \, 
\left[ \sum_{\nu'}\iint dx'dy'x'^2y'^2 {\mathcal N}^{-1/2}(x,y,x^\prime,y^\prime){\tilde \chi}_{\nu'}(\rho',\alpha')\right] 
{\mathcal A}_\nu\, |\Phi^{J^\pi T}_{\nu x y} \rangle  \,, 
\label{eq:proj_3B} 
\end{align}  
\end{widetext}
where $|\Psi_{3B}^{J^\pi T}\rangle$ is the projected wave function and 
the expression enclosed by the square brackets represents the coefficients of the expansion which are analogous to the amplitudes
$G_\nu^{J^\pi T}$ of Eq.~\eqref{eq:trialwf}. The coefficients ${\tilde \chi_\nu}$ (analogous to $\chi_\nu$ within the cluster 
part of the basis) can be calculated through the projection:  
\begin{align} 
   \tilde \chi_\nu(\rho,\alpha)=
	\langle\Psi^{J^{\pi}T}| {\mathcal A}_{\nu}|\Phi^{J^\pi T}_{\nu^\prime x^\prime y^\prime} \rangle
\label{eq:projection}    
\end{align}
where $|\Psi^{J^{\pi}T}\rangle$ is the full NCSMC wave function.  
Then, the probability density can be obtained by using $|\Psi_{3B}^{J^\pi T}\rangle$
in Eq.~\eqref{eq:prob} and reduces to 
\begin{equation}
 P(x,y) \sim x^2y^2 \sum_{\nu}\tilde\chi_{\nu}^2(x,y),    
\end{equation}
which can be expressed in terms of the NCSMC 
wave function coefficients   
$c_\lambda$ and $\chi_\nu(x,y)$ (related to $G_\nu(x,y)$ through Eq.~\eqref{eq:g_chi})    
by substituting Eq.~\eqref{eq:trialwf} in Eq.\eqref{eq:projection} when calculating $\tilde\chi_{\nu}^2(x,y)$,
i.e.,

\begin{eqnarray}
P(x,y) &\sim& x^2y^2 \sum_{\nu}\Big[ \chi_{\nu}(x,y)^2 \nonumber
\\ &+& \sum_{\lambda \lambda'}c_{\lambda}c_{\lambda'}\bar{g}_{\lambda\nu}(x,y)
\bar{g}_{\lambda'\nu}(x,y) \nonumber \\ 
&+& 2\sum_{\lambda}c_{\lambda}\bar{g}_{\lambda\nu}(x,y)\chi_{\nu}(x,y)\Big]. 
\label{eq:prob_dist}
\end{eqnarray} 

In order to have a more physical idea of the relative positions of the clusters,
the probability distribution is typically 
plotted in terms of relative distances instead
of Jacobi coordinates. 

The level of approximation within Eq.~\eqref{eq:prob_dist} can be estimated by 
calculating the integral of the probability density. Given that the wave function
is normalized, the deviation of such integral from unity represents the
part of the wave function that is not taken into account within this 
approximation.

\subsection{Radii}
\label{sec:radius}
Root-mean square matter and point-proton radii are essential observables in studying 
the spatial extension
and how inhomogeneous is their distribution of protons and neutrons.   
In general, the matter radius operator is defined as
\begin{equation}
r_m^2\equiv \frac{1}{A}\sum_{i=1}^A  
(\vec r_i-\vec R_{cm})^2,   
\label{rad_gen}
\end{equation}
where  $R_{cm}$ is the c.m. of the system, 
then the rms matter radius is given by the the square root of its expectation value.
However, for a three-cluster system, such as $^6$He, 
it can be decomposed into a relative part, which 
depends on the relative distance among the clusters and an internal part that acts on their inner coordinates.
In particular, when two of the clusters are single nucleons, the operator can be written as
\begin{equation}
r_m^2=\frac{1}{A} \rho^2+\frac{A-2}{A} r_{m}^{2(c)},   
\label{rad_3B}
\end{equation}
where $r_{m}^{2(c)}$
is the rms matter radius operator of the $A-2$-nucleon {\it core}. 

When calculating the rms matter radius within the NCSMC it is convenient to use both forms of
the operator. Indeed, while for the discrete part of the basis using the general expression~\eqref{rad_gen} 
is more appropriate, it is natural to use the cluster decomposition 
of~\eqref{rad_3B} when the three-cluster part of the basis is involved.
   
In the case of the point-proton radius we can attempt a similar cluster decomposition. 
While in this case it is not possible to obtain a simple general expression analogous to~\eqref{rad_3B},  
for the particular case in which the {\it core} 
is the only cluster with electric charge
and it is an isospin
zero state, the point-proton radius can be reduced to: 
\begin{equation}
 r_{pp}^2 \equiv \frac{1}{Z}\sum_{i=1}^A (\vec r_i-\vec R_{cm})^2 
\frac{(1+\tau^{(z)}_i)}{2}= r_{pp}^{2(c)}+ R^{2(c)}           
\label{eq:rpp}
\end{equation}
where $Z$ is the total number of protons, $r_{pp}^{(c)}$ is the rms point-proton radius operator of the {\it core} and 
$R^{(c)}=\sqrt{\tfrac{2}{A(A-2)}}\eta_{c,nn}$ is the distance
between the c.m. of the {\it core} and that of the whole system.
Similar to the matter radius, to calculate the expectation value on the NCSMC wave function, 
 the general definition of the operator (given by the central part
of Eq.~\eqref{eq:rpp}) is used when dealing with the composite part of the basis while the reduced 
form on the right of~\eqref{eq:rpp} is used when the cluster basis is involved.  

The specific expressions for the expectation values of these operators when using NCSMC wave functions
can be found in Appendix~\ref{ap:wf}.

\section{\label{sec:application} Application to $^6$He}

The $^6$He nucleus  is a prominent example of  
Borromean quantum `halo', i.e.\ a weakly-bound state of three particles 
($\alpha$+$n$+$n$) otherwise unbound in pairs, characterized by
``large probability of configurations within classically forbidden
regions of space" \cite{RevModPhys.76.215}. In the last few years, its
binding energy~\cite{PhysRevLett.108.052504} and charge
radius~\cite{PhysRevLett.93.142501} have been experimentally
determined with high precision, providing stringent tests 
for {\em ab initio} theories, including the NCSMC approach for
three-cluster dynamics presented in this paper. Further, 
the $\beta$-decay properties of the ground state (g.s.)\ of $^6$He are
of great interest for tests of fundamental interactions and
symmetries.  Precision measurements of the half life have recently 
taken place~\cite{PhysRevLett.108.122502} and efforts are under way to determine the angular 
correlation between the emitted electron and neutrino
\cite{garcia_private}.  

Less clear is the experimental picture for the low-lying continuum of
$^6$He. Aside from a narrow resonance characterized by spin-parity
$J^\pi=2^+$, 
located at 1.8 MeV above the g.s., the positions, spins and parities of the excited states of this nucleus are still under discussion. 
Resonant-like structures around $4$~\cite{PhysRevLett.85.262} and
5.6~\cite{PhysRevC.54.1070} MeV of widths $\Gamma\sim4$ and $10.9$
MeV, respectively, as well as a broad asymmetric bump at $\sim5$
MeV~\cite{Nakamura2000209}, were observed in the production of excited $^6$He
through charge-exchange reactions between two fast colliding nuclei. However,
there was disagreement on the nature of the underlying $^6$He excited
state(s).
On one hand, in Refs.~\cite{PhysRevLett.85.262}
and \cite{Nakamura2000209} these structures were attributed to 
 dipole excitations 
 compatible
 with oscillations of the positively-charged $^4$He core against the
 halo neutrons. On the other hand, 
the resonant structure of Ref.~\cite{PhysRevC.54.1070} was identified
as a second $2^+$ state. 
More recently, a much narrower $2^+$ ($\Gamma=1.6$ MeV) state at $2.6$ 
MeV as well
as a $J=1$ resonance ($\Gamma\sim2$ MeV) of unassigned parity
at $5.3$ MeV were populated with the two-neutron transfer reaction
$^8$He($p,^3$H)$^6$He$^*$~\cite{Mougeot:2012aq} at the SPIRAL facility
in GANIL.
More in general, the low-lying $\alpha$+$n$+$n$ continuum plays a
central role in the $^4$He$(2n,\gamma)^6$He radiative capture (one of the mechanism by which stars can overcome the instability of the five- and eight-nucleon systems and create heavier nuclei~\cite{doi:10.1146/annurev.nucl.48.1.175}) and of the $^3$H$(^3$H$,2n)^4$He reaction, which contributes 
to the neutron yield in fusion
experiments~\cite{PhysRevLett.109.025003,PhysRevLett.111.052501}. It
is also an important input in the evaluation of nuclear data, e.g., the $^9$Be$(n,2n)$ cross
section used in simulations of nuclear heating and material damages
for reactor technologies.

On the theory side, $^6$He has been the subject of many
investigations (see, e.g., the overviews of  Refs.~\cite{Quaglioni2013}
and \cite{Romero-Redondo:2014fya} and references therein). 
Limiting ourselves to {\em ab initio} theory, for the
most part the g.s.\ properties and low-lying excited spectrum of $^6$He have been studied within
bound-state methods, 
based on expansions on six-nucleon basis
states
~\cite{PhysRevC.65.054302,Pieper2008,PhysRevC.68.034305,PhysRevC.73.021302,Bacca:2012up,Saaf2014,Caprio2014,Constantinou2016}.
These include: Monte Carlo~\cite{PhysRevC.65.054302,Pieper2008} and NCSM~\cite{PhysRevC.68.034305}
calculations of the g.s.\ energy, point-proton radius, $\beta$-decay
transition and excitation energies based on $NN+3N$ interactions; 
a large-scale NCSM study of the matter and point-proton radii with $NN$
interactions~\cite{PhysRevC.73.021302}; a
hyperspherical harmonics
study of the correlation between two-neutron separation energy and the
matter and charge radii using low-momentum $NN$
potentials~\cite{Bacca:2012up};  an investigation of the
$\alpha$+$n$+$n$ channel
form factors of NCSM g.s.\ solutions obtained with soft $NN$
interactions and (in a more limited space) 3N forces~\cite{Saaf2014};
and no-core configuration interaction calculations within a Coulomb
Sturmian~\cite{Caprio2014} and natural orbital~\cite{Constantinou2016}
basis, starting from the JISP16 $NN$ interaction.
In general, these {\em ab initio} calculations
describe successfully the interior of the $^6$He wave function, 
but are unable to fully account for its three-cluster 
asymptotic behavior. As a consequence, the
simultaneous reproduction of the small binding energy and extended
radii of $^6$He has been a challenge. Further, the low-lying resonances
of $^6$He have been treated as bound states, an approximation that is
well justified only for the narrow $2^+$ first excited state, and that
does not provide
information about their widths.

An initial description of $\alpha$+$n$+$n$ dynamics within an {\em ab
  initio} framework was achieved using a soft $NN$ potential in our earlier studies of
Refs.~\cite{Quaglioni2013} and \cite{Romero-Redondo:2014fya}, carried out 
in a model space spanned only by $^4$He(g.s.)+$n$+$n$
continuous basis states of the type of Eq.~\eqref{eq:3bchannel}. This approach 
naturally explained the asymptotic configurations of the $^6$He
g.s.\ and enabled the description of $\alpha$+$n$+$n$ continuum
states, but was unable to fully account for
short-range many-body correlations, as clearly indicated by the
underestimation of the g.s.\ energy. This shortcoming was later
addressed in Ref.~\cite{rom2016}, where we achieved a simultaneous
description of six-body correlations
and $\alpha$+$n$+$n$ dynamics working within the framework of the
three-cluster NCSMC, presented in this paper.  

In the following we
discuss the calculations of Ref.~\cite{rom2016}, as well as additional results, more extensively.
The adopted NCSMC model space
includes the first nine positive-parity, and first six negative-parity   
 square-integrable eigenstates of $^6$He with $J\le2$, 
obtained by diagonalizing the Hamiltonian within the
six-body HO basis of the NCSM,  
as well as $^4$He(g.s.)+$n$+$n$ three-cluster 
channels for which the $^4$He core is also described within the NCSM.
Calculations are performed using the chiral N$^3$LO $NN$ potential or
Ref.~\cite{N3LO} softened via the similarity renormalization group (SRG)
method~\cite{PhysRevC.75.061001,
PhysRevC.77.064003,Wegner1994}, and disregard for the time being $3N$
initial and SRG-induced components of the nuclear Hamiltonian. This
defines a new $NN$ interaction, denoted SRG-N$^3$LO $NN$, unitarily
equivalent to the initial potential in the two-nucleon sector only. 
Specifically, we adopt the resolution-scale parameters $\lambda_{\rm SRG}=1.5$ fm$^{-1}$
and $\lambda_{\rm SRG}=2.0$ fm$^{-1}$, and the same $\hbar\Omega=$14 and 20 MeV HO frequencies used in Refs. 
\cite{Quaglioni2013,Romero-Redondo:2014fya} and \cite{PhysRevLett.114.212502}, 
respectively. The results obtained with the   
$\lambda_{\rm SRG}=1.5$ fm$^{-1}$ resolution scale provide a benchmark
for the method given that, with such a  soft potential, reliable
values for the g.s.\ and $2^+_1$  
energies can be extracted, by extrapolation to the `infinite' space,
from a NCSM calculation. 
Furthermore, the results obtained with this potential can be directly compared with
those of Refs.~\cite{Quaglioni2013,Romero-Redondo:2014fya},
using expansions based exclusively on $^4$He(g.s.)+$n$+$n$ microscopic
cluster states. 
Such comparison allows us to better understand the importance of the short
range correlations that were missing in that calculation.
Conversely, calculations carried out with the $\lambda_{\rm SRG}=2.0$
fm$^{-1}$ resolution scale 
allow for a `more realistic' study of the g.s.\ properties
of $^6$He. Indeed,  at this momentum scale the net effects of the 
disregarded initial and SRG-induced 3$N$ interaction is mostly suppressed in nuclei up to mass number $A=6$, leading to   
binding energies close to
experiment~\cite{PhysRevC.83.034301}. Furthermore, for this resolution scale
two- and higher-body SRG corrections to the $^3$H and $^4$He matter radii
computed with bare operators (as done in the present work) 
have been shown to be negligible (less than $1\%$)~\cite{Schuster2014}. 
\begin{figure}[ht]
      \includegraphics[width=8.5cm,clip=,draft=false]{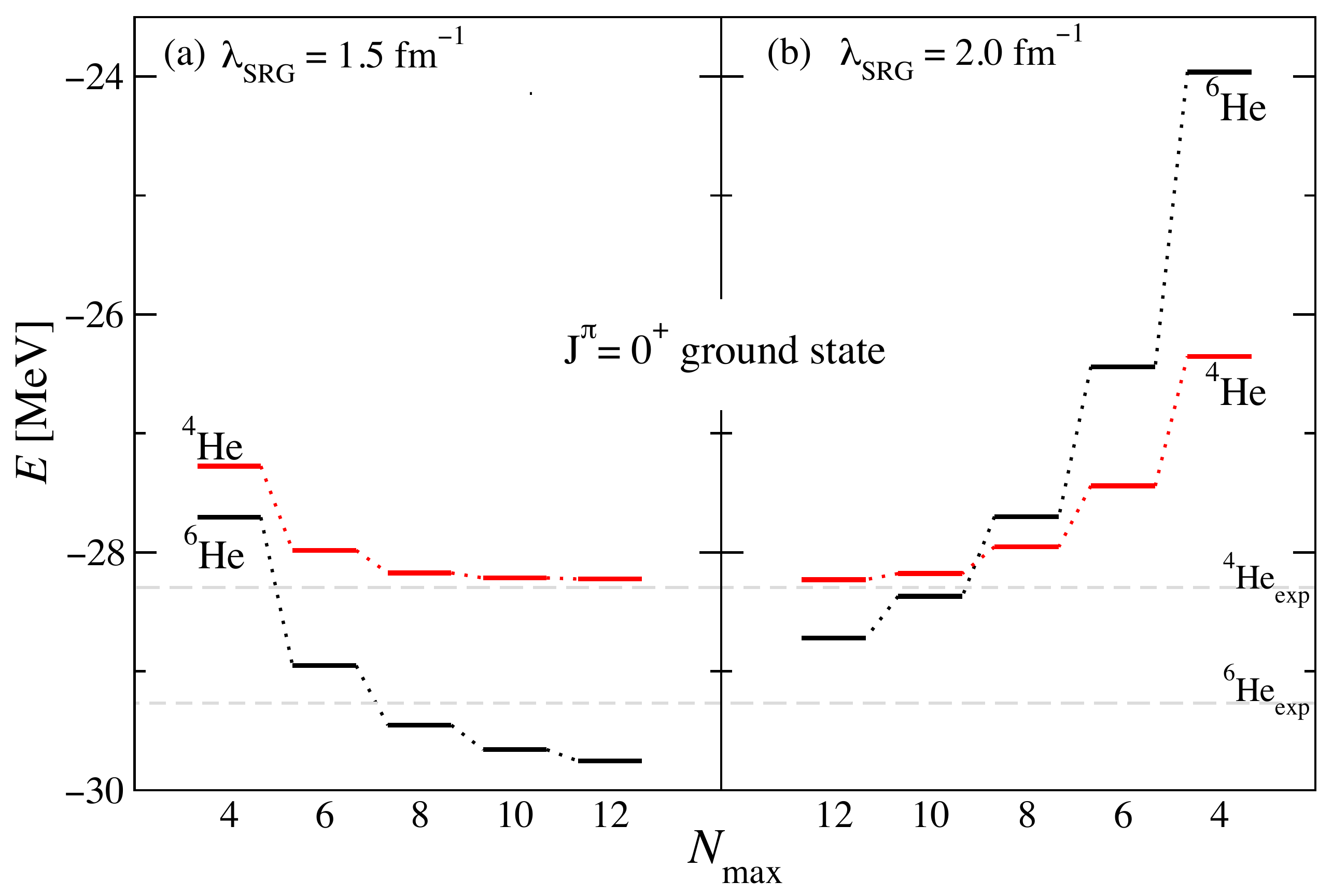}  
        \caption{Dependence of the NCSM $^6$He and $^4$He $J^\pi=0^+$ ground state energies $E$(g.s.) on the HO model space size $N_{\rm max}$ for the SRG-N$^3$LO $NN$ potential with (a) $\lambda_{\rm SRG}=1.5$ fm$^{-1}$ and $\hbar\Omega=14$ MeV, and (b) $\lambda_{\rm SRG}=2.0$  fm$^{-1}$ and $\hbar\Omega=20$ MeV.} 
\label{NCSM_0p}
\end{figure}  

\subsection{$^4$He and $^6$He square integrable eigenstates}  
\label{sec:NCSM}
\begin{table}[t]
\caption{Energy ($E_x$) of the first 9 positive-parity states with $J\le2$ for $^6$He calculated within the NCSM 
for a model space of $N_{\mbox{max}}$=12.} 
\begin{ruledtabular}
\begin{tabular}{c c c}  
$J^\pi$ &$\lambda_{\rm SRG}= 1.5$ fm$^{-1}$  &  $\lambda_{\rm SRG}= 2.0$ fm$^{-1}$\\
\hline
0$^+$ & $-29.75$ & $-28.72$ \\
      & $-22.73$ & $-20.10$\\
      & $-20.46$ & $-15.25$\\
      & $-19.04$ & $-13.39$\\
1$^+$ & $-24.25$ &  $-22.28$\\ 
      & $-18.77$ & $-13.57$\\
2$^+$ & $-27.40$ &  $-26.24$\\
      & $-24.78$ &  $-22.99$\\
      & $-19.22$ & $-13.84$\\
%3$^+$ & -18.897& -13.733\\
\end{tabular}
\end{ruledtabular}
\label{pos_par}
\end{table}
\begin{table}[b]
\caption{Energy ($E_x$) of the first 6 negative-parity states with $J\le2$ for $^6$He calculated within the NCSM  
for a model space of $N_{\mbox{max}}$=13.} 
\begin{ruledtabular}
\begin{tabular}{c c c}  
$J^\pi$ &$\lambda_{\rm SRG}= 1.5$ fm$^{-1}$  &  $\lambda_{\rm SRG}= 2.0$ fm$^{-1}$\\
\hline
0$^-$ & $-21.40$ & $-17.84$\\
1$^-$ & $-23.84$ & $-20.97$ \\
      & $-21.63$ & $-17.98$\\
      & $-19.90$ & $-16.12$\\
2$^-$ & $-23.33$ & $-20.45$\\
      & $-19.67$ & $-15.96$\\ 
\end{tabular}
\end{ruledtabular}
\label{neg_par}
\end{table}
\begin{figure*}[t]
      \includegraphics[width=14cm,clip=,draft=false]{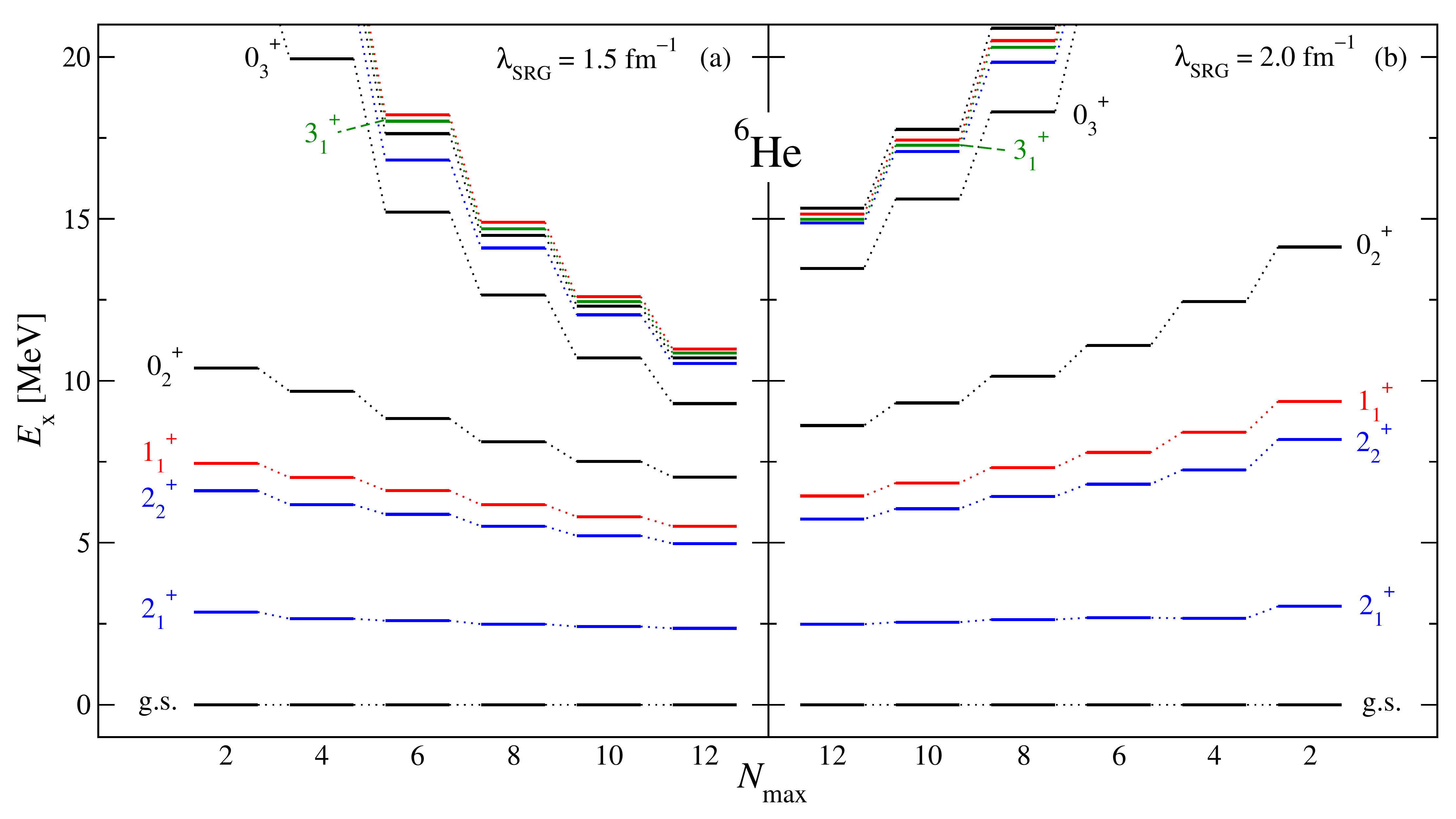} 
        \caption{Dependence of the NCSM $^6$He $J^\pi=0^+$ excitation energies ($E_{\rm x}$) on the HO model space size $N_{\rm max}$ for the SRG-N$^3$LO $NN$ potential with (a) $\lambda_{\rm SRG}=1.5$ fm$^{-1}$ and $\hbar\Omega=14$ MeV, and (b) $\lambda_{\rm SRG}=2.0$  fm$^{-1}$ and $\hbar\Omega=20$ MeV.} 
\label{NCSM_all_exc}
\end{figure*}  
In this section, we discuss our results for the NCSM
eigenstates used as input for the present NCSMC investigation of the
$J^\pi=0^+$ g.s.\ of $^6$He and low-lying $\alpha$+$n$+$n$ continuum for partial waves up to $J^\pi=2^{\pm}$.

The computed energy of the $^6$He g.s.\ within the NCSM is presented in Fig.~\ref{NCSM_0p}
as a function of the HO basis size $N_{\rm max}$.
Results obtained with
$\lambda_{\rm SRG}=1.5$ fm$^{-1}$ and $\hbar\Omega=14$ MeV, shown in
panel (a), are compared with those in panel (b) for  $\lambda_{\rm SRG}=2.0$  fm$^{-1}$ and $\hbar\Omega=20$ MeV.
For the softer ($\lambda_{\rm SRG}=1.5$ fm$^{-1}$) potential, the variational NCSM
calculations converge rapidly and can be easily
extrapolated to $N_{\rm max}\rightarrow\infty$ using an
exponential function of the type 
\begin{equation}
E(N_{\rm max})=E_{\infty}+a e^{-b 
  N_{\rm max}}.   
\label{expo_fit}
\end{equation}
This yields $E$(g.s.)$=
-29.84(4)$~\cite{Quaglioni2013}, which is about 0.6 MeV overbound with respect to
experiment. 
The convergence rate is clearly slower for the $\lambda_{\rm SRG}=2.0$ fm$^{-1}$
interaction.
Nevertheless, also in this case, the infinite-space g.s.\
energy can be accurately obtained using the extrapolation techniques
recently developed for the
NCSM~\cite{Coon2012,Furnstahl2012,More2013,Furnstahl2014,Wendt2015}.
This was recently demonstrated by S\"a\"af and
Forss\'en, who obtained the extrapolated value of $E$(g.s.)$=  
-29.20(11)$ MeV~\cite{Saaf2014} in close agreement with experiment 
(-29.268 MeV).
Also shown in Fig.~\ref{NCSM_0p} are the corresponding results for the energy of the $^4$He g.s.,
which is used to build the microscopic cluster states of Eq.~\eqref{eq:3bchannel}.
For both $\lambda_{\rm SRG}$ values convergence is achieved within the largest HO model space, yielding binding energies close to experiment, 
as was already shown in Ref.~\cite{PhysRevC.83.034301}.
 
Figure~\ref{NCSM_all_exc} shows the convergence pattern with respect to the HO basis size of the excitation energies for the first 10 positive-parity NCSM eigenstates of $^6$He. These include four 0$^+$ , two 1$^+$ and 
three 2$^+$ states, and one $3^+$ state.  This latter state is not used in the present NCSMC calculations.
As before, the results obtained with the $\lambda_{\rm SRG}= 1.5$ and $2.0$ fm$^{-1}$ interactions are shown in panel (a) and (b), respectively.  
Except for the $2^+_1$ state, which presents a very mild $N_{\rm max}$ dependence, the convergence rate is steady but slow, and tends to deteriorate as the excitation energy increases. The convergence rate is once again much faster for the softer potential, which also generates a more compressed excitation spectrum compared to the $\lambda_{\rm SRG}=2.0$ fm$^{-1}$ interaction. 
The overall picture is similar for the negative-parity states. A summary of the NCSM eigenenergies used as input in the largest model space adopted is given
in Tables~\ref{pos_par} and \ref{neg_par} for positive and negative parities, respectively.

\begin{figure}[b]
      \includegraphics[width=8.5cm,clip=,draft=false]{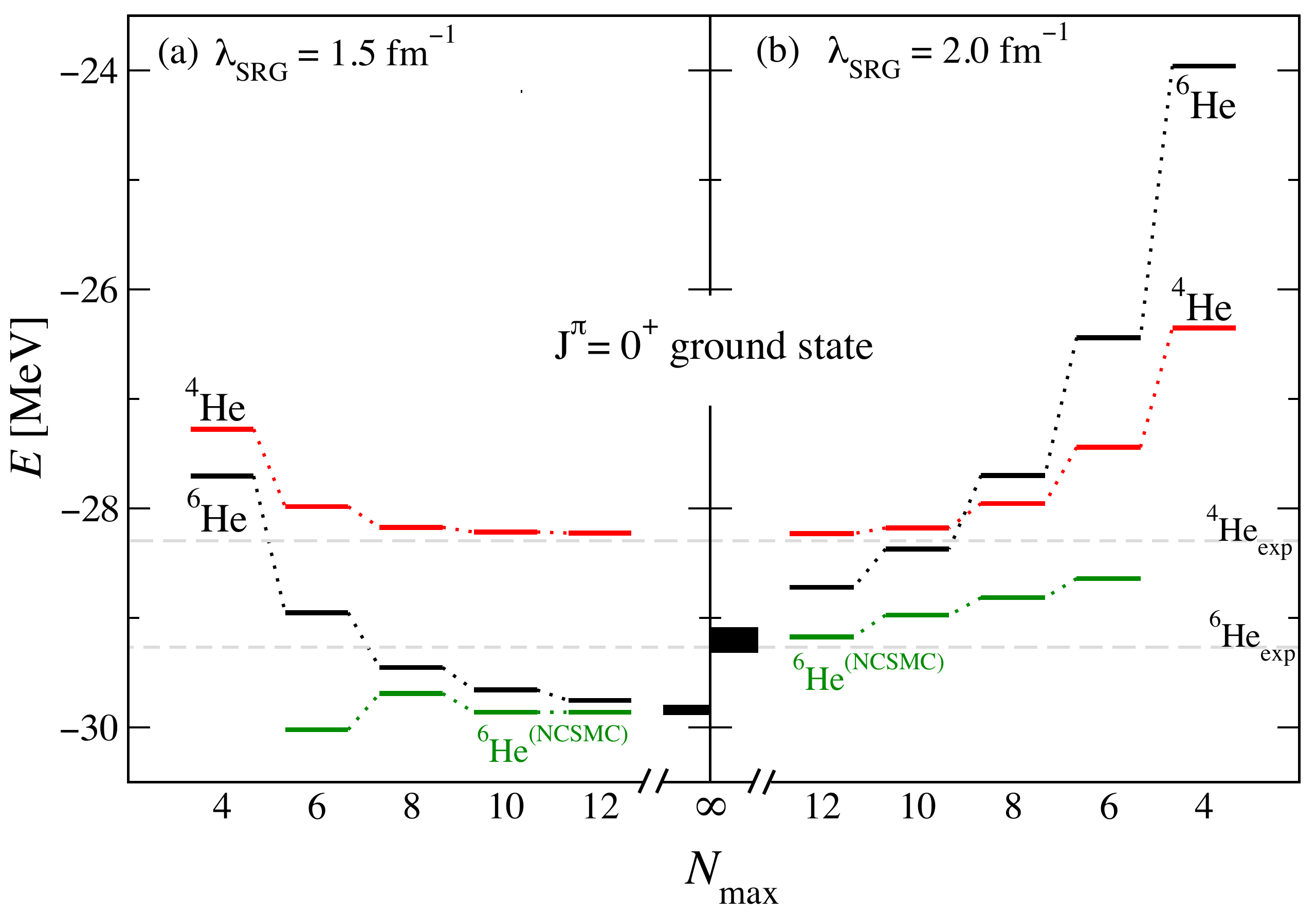}  
        \caption{Same as Fig.~\ref{NCSM_0p} including also the dependence of the NCSMC $^6$He $J^\pi=0^+$  ground state energy ($E$) on the 
HO model space size $N_{\rm max}$ for the SRG-evolved N$^3$LO $NN$ potential with (a) $\lambda_{\rm SRG}=1.5$ fm$^{-1}$ and 
$\hbar\Omega=14$ MeV, and (b) $\lambda_{\rm SRG}=2.0$  fm$^{-1}$ and $\hbar\Omega=20$ MeV. The extrapolated $N_{\rm max}\to \infty$ NCSM $^6$He  
is shown as a band of which the width represents the extrapolation uncertainty.}    
\label{NCSMC_0p}
\end{figure}  
\begin{table*}
\caption{Computed $^6$He g.s.\ energies (in MeV) within the cluster basis  
[$^4$He(g.s.)+$n$+$n$]~\cite{Quaglioni2013} ($4^{\rm th}$ column), NCSM ($5^{\rm th}$ column) and 
NCSMC  including $N_\lambda=1$ eigenstate of the composite system ($6^{\rm th}$ column) as a function of the HO model space size $N_{\rm max}$
for the SRG-evolved N$^3$LO $NN$ potential with $\lambda_{\rm SRG}=1.5$ fm$^{-1}$.  Also shown for the biggest model space are the results for the NCSMC including $N_\lambda=4$  $^6$He eigenstates, and the NCSM $^6$He energy obtained through the exponential fit from Eq.~\eqref{expo_fit}. Results for $^4$He and experimental values for $^6$He  are presented in the $3^{\rm rd}$ and last column, respectively.}
\begin{ruledtabular}
\begin{tabular}{c c c c c c c}
$N_{\rm max}$ & ($N_\lambda$) & $^4$He NCSM &$^6$He \cite{Quaglioni2013} & $^6$He NCSM & $^6$He NCSMC & $^6$He Expt.\\   
\hline
6  & (1) &   $-27.98$ &$-28.91$ & $-28.95$ & $-30.02$&\\
8  & (1) &   $-28.17$ &$-28.62$ & $ -29.45$ & $-29.69$&\\
10& (1) &  $-28.21$ &$-28.72$ & $-29.66$ & $-29.86$&\\
12& (1) &  $-28.22$ &$-28.70$ & $-29.75$ & $-29.86$&  $-29.268$~\cite{PhysRevLett.108.052504}\\ 
12& (4) &--&--&--& $-29.88$& \\
14&  -- & $-28.22$ &-- & --&--&\\
$\infty$ & --& --&  -- & $-29.84(4)$ & -- &\\
\end{tabular}
\end{ruledtabular}
\label{energy}
\end{table*}

\subsection{$^6$He ground state within the NCSMC}

\begin{table}[b]
\caption{Same as Table~\ref{energy}, now using the potential obtained with
a SRG evolution parameter of $\lambda=$2.0 fm$^{-1}$. 
 The NCSM extrapolation shown is the one from Ref. \cite{Saaf2014}.  
 Note that for this potential the cluster basis alone does not yield
a bound $^6$He ground state. } 
\begin{ruledtabular}
\begin{tabular}{c c c c c }
$N_{\rm max}$ & ($N_\lambda$)& $^4$He NCSM & $^6$He NCSM & $^6$He NCSMC\\   
\hline
6   &  (1) &$-27.44$  & $-26.44$ & $-28.31 $\\
8   &  (1) &$-27.95$  & $-27.70$ & $-28.81$\\
10 &  (1) &$-28.18$  & $-28.37$ & $-28.97$\\
12 &  (1) &$-28.23$  & $-28.72$ & $-29.17$\\ 
12 &  (4) &--&--& $-29.17$\\
14 &   --  & $-28.24$ &-- & --\\
$\infty$ & --  & -- & $-29.20(11)$\cite{Saaf2014}  \\
\hline\\[-2.5mm]
\multicolumn{2}{c}{$^6$He Expt.} &\multicolumn{3}{c}{  $-29.268$~\cite{PhysRevLett.108.052504}}\\
\end{tabular}
\end{ruledtabular}
\label{energy2}
\end{table}
The convergence of the $^6$He g.s.\ energy computed within the NCSMC in terms of the size of the model space is 
compared with the corresponding NCSM results in Fig.~\ref{NCSMC_0p}. 
More detailed comparisons (including with the results obtained working in a cluster basis alone \cite{Quaglioni2013}) 
are presented in Tables~\ref{energy} and \ref{energy2} for the $\lambda_{\rm SRG}=1.5$ fm$^{-1}$ and $\lambda_{\rm SRG}=2.0$ fm$^{-1}$
interactions, respectively.
The $3^{\rm rd}$ column of Table~\ref{energy} shows 
the energy of the ground state of $^4$He within the NCSM, which defines the three-body
breakup energy threshold $E_{th}(\alpha+n+n)$ for all present $^6$He calculations. 
This is clearly already converged at the largest adopted model space size.  
The next three columns show the energy of the g.s.\ of $^6$He calculated within the
$^4$He(g.s.)+$n$+$n$ cluster basis of Ref.~\cite{Quaglioni2013}, the NCSM and NCSMC. We can see that 
the fastest convergence is reached within the NCSMC. Furthermore, while the results from 
Ref.~\cite{Quaglioni2013} also present a weak dependence on the HO model space size,
they do not converge to the correct energy, which can be estimated by extrapolating to the infinity model space
the NCSM results. This proves that the many-body correlations
disregarded when using the cluster basis alone are indeed necessary for the correct 
description of the system and are correctly taken into account within the
NCSMC. 
While the convergence of the NCSMC $^6$He g.s.\ energy with respect to the model space size is shown here for the case in which only  one eigenstate of the composite system is included in the calculations, we also present the result obtained by including four eigenstates of $^6$He for the largest model
space size. This shows that  the   
inclusion of additional eigenstates of the composite system has only a small effect on the g.s.\ energy.

It is worth noting that the NCSMC 
is a variational approach as long as the adopted model space captures in full the wave function 
of the clusters (here, the $^4$He core) and of the aggregate system (here, $^6$He) or, equivalently, 
if it includes all possible pre-diagonalized eigenvectors of the clusters and of the aggregate 
system within the chosen $N_{\rm max}$ HO basis size. That is,  the NCSMC 
is a variational approach as long as the generalized cluster 
expansion is not truncated. Such a model space is computationally unachievable and, for p-shell nuclei,   
we truncate the generalized cluster expansion to include only a few eigenstates of the cluster 
and aggregate nuclei.  In particular, in the present application we only include the g.s.\ of the $^4$He core. 
The effect of this truncation manifest itself in the smallest HO base sizes, 
and can give rise to the non-variational behavior shown in Table~\ref{energy} (the same argument applies to the
cluster basis calculation of Ref.~\cite{Quaglioni2013}). However, as the adopted HO basis size increases, 
thanks to the overcomplete nature of the NCSMC basis the wave functions of clusters and aggregate 
system are better and better represented within the truncated cluster expansion and the convergence 
behavior becomes variational, with the typical approach to the g.s.\ energy from above. 

In Ref. \cite{rom2016} the equivalent results were presented in terms of the absolute HO model space size 
$N_{\rm tot}=N_{0}+N_{\rm max}$, where $N_0$ 
is the number of quanta shared by the nucleons in their lowest configuration. However, given  
 that the input for the NCSMC includes the elements of the composite and cluster bases at 
the same $N_{\rm max}$, we came to the conclusion that a comparison in terms of $N_{\rm max}$ provides a 
better picture of the relevance of each component
in the full calculation. 
We also note that the last three columns of  
Table I in Ref.~\cite{rom2016} present
 a mismatch with respect to the model space size reported in the first column, 
showing results obtained with an $N_{\rm{max}}$ value larger by 2 units.
Therefore, we call the reader to consider the present tables to be 
the accurate representation of the results. 

\begin{table}[t]
\caption{Percentage of the norm of the $^6$He g.s.\ wave function that comes directly from the  
NCSM part of the basis ($\sum_{\lambda}c_{\lambda}^2$).}
\begin{ruledtabular}
\begin{tabular}{c c c} 
$N_{\rm max}$   & $\lambda_{\mbox{\scriptsize{SRG}}}=$1.5 fm$^{-1}$ & $\lambda_{\mbox{\scriptsize{SRG}}}=$2.0 fm$^{-1}$ \\ [1mm] 
\hline\\[-2.5mm]
8&    $78\%$ & ---\\
10&   $88\%$ & $71\%$\\
12&   $91\%$ & $76\%$  \\
\end{tabular}
\end{ruledtabular}
\label{composition}
\end{table}
\begin{figure}[b]
      \includegraphics[width=7cm,clip=,draft=false]{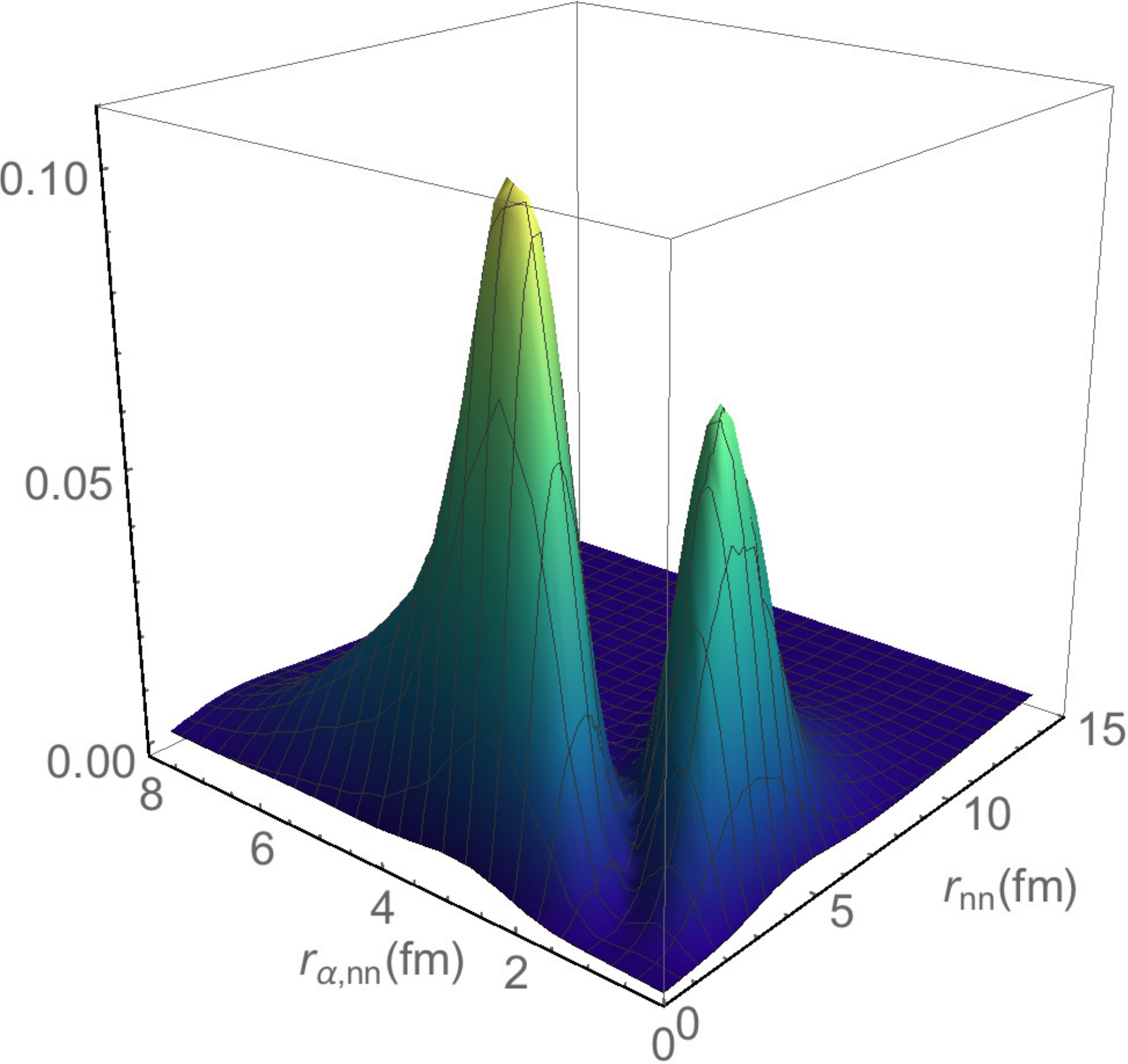} 
\caption{(Color online) Probability distribution the 
$J^{\pi}=0^+$ ground state of the $^6$He. 
Here $r_{nn}=\sqrt{2}\,\eta_{nn}$ and $r_{\alpha,nn}=\sqrt{3/4}\,\eta_{\alpha,nn}$ 
are, respectively, the distance between the two neutrons and the distance between the c.m. of $^4$He and that of the two neutrons.}
\label{prob_dist}
\end{figure}
\begin{figure*}[t]
\includegraphics[width=5.2cm,clip=,draft=false]{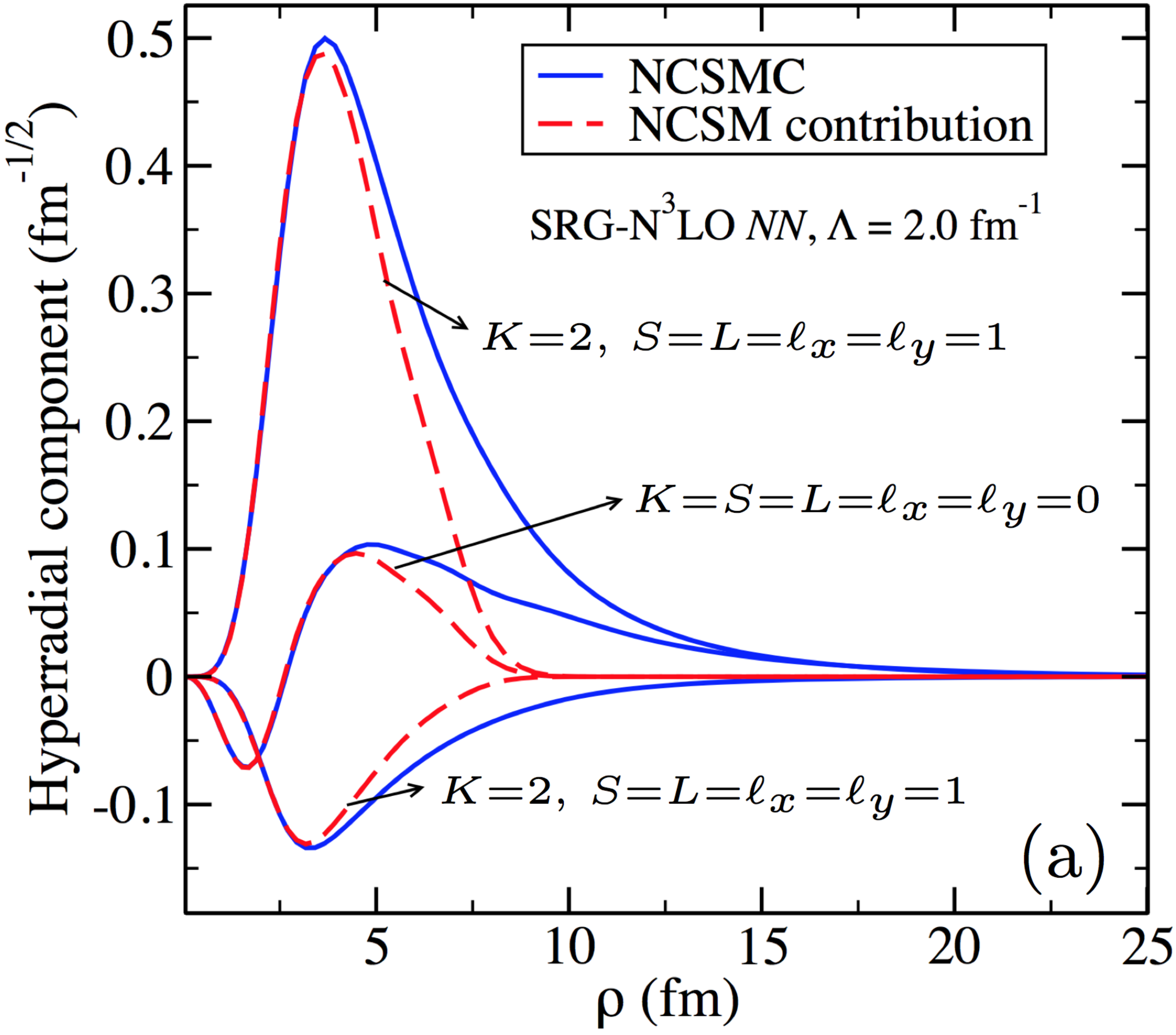}
\qquad
      \includegraphics[width=5.5cm,clip=,draft=false]{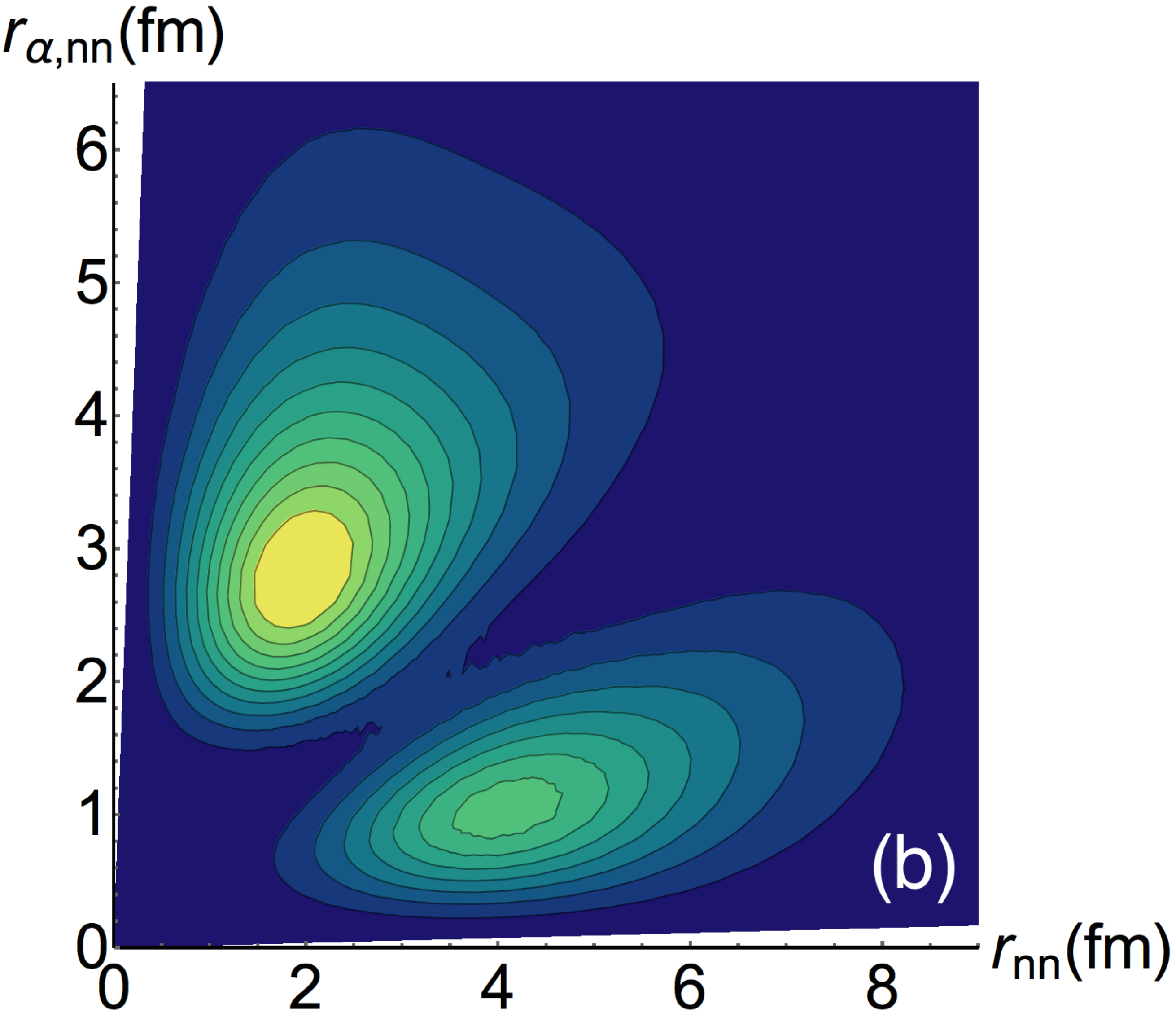} 
      \includegraphics[width=5.5cm,clip=,draft=false]{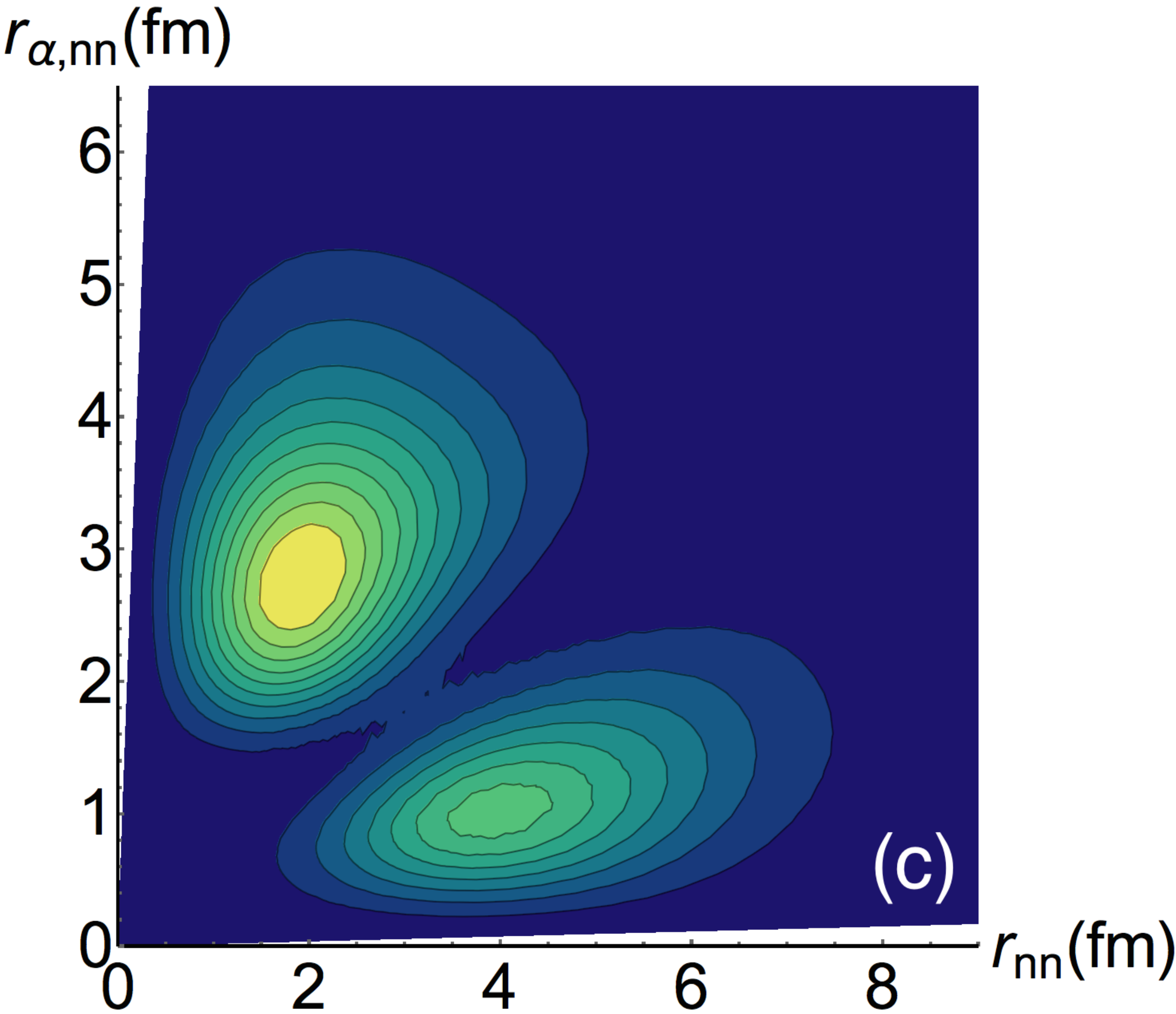} 
\caption{(Color online) Panel (a): Most relevant hyperradial components $\tilde u_{\nu K}(\rho)$ [see text] of the $\alpha$+$n$+$n$ relative motion
 within the $^6$He g.s.\ after projection of the $\lambda=2.0$ fm$^{-1}$ 
full NCSMC wave function in the largest model space (blue solid lines) as well as of its NCSM portion (red dashed lines) 
into the orthogonalized microscopic-cluster basis. 
Panel (b) and (c): Contour plots of the probability distribution 
obtained from the projection of the full NCSMC wave function of panel (a) and its NCSM component, respectively, 
as a function of the relative coordinates $r_{nn}=\sqrt{2}\,\eta_{nn}$ and $r_{\alpha,nn}=\sqrt{3/4}\,\eta_{\alpha,nn}$.
}     
\label{contour}
\end{figure*}

As seen in Table~\ref{energy2}, 
convergence is not as obviously reached  when using the harder potentials
with  $\lambda_{\mbox{\scriptsize{SRG}}}=2.0$ fm$^{-1}$. Within the NCSMC, there still 
are 200 keV difference between the $N_{\rm max}$= 10 and 12 results. However, the fact that the
value obtained for $N_{\rm max}$= 12 (-29.17 MeV) is in agreement with the    
 NCSM extrapolation from Ref.~\cite{Saaf2014} (-29.20(11)) is a good indicator that
our results are at least very close to convergence at this model space size.

We can estimate how much of the wave function can be described through the NCSM by
calculating the percentage of the norm that comes directly from the discrete part of the basis,
i.e. $\sum_{\lambda} c_{\lambda}^2$. These percentages are shown in Table~\ref{composition}
for the two different potentials used, as well as for different sizes of the model space.
We find that, as one would expect, the NCSM component of the basis is able to describe a much larger percentage of the wave function 
when using the softer potential corresponding to the $\lambda_{\rm{SRG}}=1.5$~fm$^{-1}$ resolution scale,
and also a larger and larger percentage as the HO model space size increases. 

\subsubsection{Spatial distribution}

In Fig.~\ref{prob_dist} we show the probability density, as defined in 
section~\ref{sec:prob_dist},
 for the ground state of $^6$He in terms of the the distance 
between the two halo neutrons 
($r_{nn}=\sqrt{2}\,\eta_{nn}$) and the distance between the $^4$He $core$ and the 
center of mass of the external neutrons ($r_{\alpha,nn}=\sqrt{3/4}\,\eta_{\alpha,nn}$).
This density plot presents two peaks, which correspond to the two preferred spatial
configurations of the system. The di-neutron configuration, 
which corresponds to the two neutrons being close together, clearly presents a higher 
probability respect to the cigar configuration in which the two neutrons are far apart
and at the opposite sides of the $core$. This distribution is in agreement 
with previous studies \cite{Descouvemont:2003ys,
Brida:2010ae,Quaglioni2013,kukulin86,Zhukov1993151,Saaf2014,Nielsen2001373}. 
In order to estimate the reliability of the approximation of Eq.~\eqref{eq:prob_dist},
which uses the projection of the NCSMC wave function into the cluster basis, we   
 integrated the probability 
density given by Eq.~\eqref{eq:prob_dist}. This integral is equivalent to the square of
the norm of the projected wave function. We obtained 0.971 for the potential 
with  $\lambda_{\rm SRG}=1.5$ fm$^{-1}$ and
0.967 for the potential with
$\lambda_{\rm SRG}=2.0$ fm$^{-1}$. Given that we work with normalized wave functions, 
the proximity of these integrals to the unity  
indicates that only a small 
part of the wave functions was lost when performing the projection. %
  
\begin{table*}[t]
\caption{Computed $^4$He and $^6$He matter radii (in fm) for $\lambda_{\rm{SRG}}=1.5$~fm$^{-1}$ and $\lambda_{\rm{SRG}}=2.0$~fm$^{-1}$ as a function of the HO model space size $N_{\rm max}$ within the NCSM, and the NCSMC including $N_\lambda=1$ eigenstates of the composite system.  Also shown for the biggest model space are the results for the NCSMC including $N_\lambda=4$  $^6$He eigenstates. Experimental values for $^6$He  are presented in the last column.}   
\begin{ruledtabular}
\begin{tabular}{c c c c c c c c c c c c}  
& & &\multicolumn{3}{c}{$\lambda_{\rm SRG}=1.5$ fm$^{-1}$} && \multicolumn{3}{c}{$\lambda_{\rm SRG}=2.0$ fm$^{-1}$}&& Expt.\\[1mm] 
%\hline
$N_{\rm max}$ &($N_\lambda$) &&  $^4$He NCSM & $^6$He NCSM& $^6$He NCSMC && $^4$He NCSM & $^6$He NCSM & $^6$He NCSMC && $^6$He \\[0.5mm] 
\hline\\[-2.5mm]
6   & (1) &&   1.489   & $2.14$ & $2.47$ && 1.471  & $2.01$ & $2.47$ && \\ 
8   & (1) &&   1.490   & $2.18$ & $2.35$ && 1.461  & $2.06$ & $2.40$&&  2.33(4)\cite{Tanihata:1992wf}\\
10 & (1) &&  1.487   & $2.22$ & $2.38$ && 1.461  & $2.10$ & $2.42$&& 2.30(7)\cite{Alkhazov:1997zz}\\
12 & (1) &&  1.490   & $2.25$ & $2.37$ && 1.459  & $2.15$ & $2.41$ &&  2.37(5)\cite{kiselev05}\\ 
%\hline
12 & (4) && -- &--&--&&--&--&2.46(2)&&
\end{tabular}
\end{ruledtabular}
\label{tab:mat_rad}
\end{table*}
\begin{table*}[t]
\caption{Computed $^4$He and $^6$He point-proton radii (in fm) for $\lambda_{\rm{SRG}}=1.5$~fm$^{-1}$ and $\lambda_{\rm{SRG}}=2.0$~fm$^{-1}$ as a function of the HO model space size $N_{\rm max}$ within the NCSM, and the NCSMC including $N_\lambda=1$ eigenstates of the composite system.  Also shown for the biggest model space are the results for the NCSMC including $N_\lambda=4$  $^6$He eigenstates. Experimental values for $^6$He  are presented in the last column.} 
\begin{ruledtabular}
\begin{tabular}{c c c c c c c c c c c c} 
& && \multicolumn{3}{c}{$\lambda_{\rm SRG}=1.5$ fm$^{-1}$} && \multicolumn{3}{c}{$\lambda_{\rm SRG}=2.0$ fm$^{-1}$}&& Expt.\\[1mm]  
$N_{\rm max}$ & ($N_\lambda$)& & $^4$He NCSM  & $^6$He NCSM & $^6$He NCSMC &&  $^4$He NCSM & $^6$He NCSM & $^6$He NCSMC && $^6$He\\[0.5mm] 
\hline\\[-2.5mm]
6  &  (1) &&   1.501 & $1.75$ & $1.92$ && 1.474 & $1.68$ & $1.91$&&\\
8  &  (1) && 1.493 & $1.77$ & $1.85$ && 1.464 & $1.70$ & $1.86$&&\\
10&  (1) && 1.490 & $1.78$ & $1.86$ && 1.464 & $1.72$ & $1.89$&& 1.938(23)~\cite{Bacca:2012up}\\
12&  (1) && 1.487 & $1.79$ & $1.85$ && 1.462 &$1.74$ & $1.87$&&\\  
12  & (4) &&--&--&--&&--&--&1.90(2) &&
\end{tabular}
\end{ruledtabular}
\label{p_radius}
\end{table*}
When the $^6$He ground state wave function is calculated
within the NCSM basis,
the probability density can be obtained by projecting into a cluster basis in the same
way as it is done for the NCSMC in Eq.~\eqref{eq:proj_3B}. The obtained projected wave function presents the same 
 distribution observed in the case of the NCSMC, with the difference that it is less extended.
 This picture is consistent with the results previously reported
 in Ref.~\cite{Saaf2014}, and is
to be expected given that within this basis the three-body asymptotic behavior is not well 
described.     
This is easily appreciated in Fig.~\ref{contour}, 
where the contour diagram of the probability distribution 
is shown for the NCSMC in panel (b) and for the NCSM component in panel (c). 
In the contour plots, it is also easier to determine the position on the
probability maxima: within the 
di-neutron configuration the highest probability density appears when the  
neutrons are about 2 fm apart and the $^4$He $core$ about 3 fm from them. 
Within the cigar configuration, the neutrons are about 4 fm apart and the
$core$ around 1 fm from their center of mass. 

In panel (a) of Fig.~\ref{contour}, the most relevant hyperradial components $\tilde{u}_{\nu K}(\rho)$
of the $\alpha$+$n$+$n$ relative motion 
are shown. The hyperradial components $\tilde{u}_{\nu K}(\rho)$ are analogous to  
${u}_{\nu K}(\rho)$ from Eq.~\eqref{eq:expansion} but defined for the projected 
wave function from Eq.~\eqref{eq:proj_3B}. 
 The solid blue lines are the components from the full NCSMC
wave function while the dashed red lines represent the contribution to  
the full NCSMC wave function coming from the discrete NCSM eigenstates. 
This figures also provides a good visualization
of how the short range of the NCSM wave function is complemented 
with the cluster basis to reproduce the extended wave function typical of
halo nuclei by means of the NCSMC.

\subsubsection{Radii}

The spatial extension of a particular state can be estimated by its matter radius
as described in section~\ref{sec:radius}. 
In table~\ref{tab:mat_rad}, we show the calculated NCSMC rms matter radius for
the  ground state of $^6$He as a function of the HO model space size $N_{\rm max}$. 
Results are shown for both $\lambda_{\rm SRG}=1.5$ fm$^{-1}$ and $\lambda_{\rm SRG}=2.0$ fm$^{-1}$. 
The results obtained within the NCSM alone  are also shown for comparison. 
The introduction of $^4$He(g.s.)+$n$+$n$ microscopic cluster 
basis states provides a matter radius
closer to experiment within smaller model
spaces.  
Contrary to the NCSM, the 
convergence of the radius with respect to the size of the model space is achieved within
the NCSMC at computationally accessible model spaces. 
 The importance of 
the inclusion of the cluster states is even more pronounced for the
potential with
$\lambda_{\rm SRG}$=2.0 fm$^{-1}$, for which the NCSM results are further away 
from convergence.
Similar to the g.s.\ energy discussed earlier, here too the convergence of the NCSMC is studied for the case in which only one eigenstate of the composite system is included in the calculation. In the largest HO model space, the inclusion of 3 additional (4 total) square-integrable eigenstates of the $^6$He system, yields a 2\% increase of the matter radius. Besides the contributions coming from the rms matter radii of the additional discrete basis states, which are largely suppressed by the fact that the corresponding expansion coefficients ($c_\lambda$) are small, such an increase comes from the matrix elements of the matter radius operator between the first and third $0^+$ square integrable basis states.
Our most complete results of $2.46(2)$ fm lies just above the range of experimental matter radii spanned by the values and associated error bars of Refs.~\cite{Tanihata:1992wf,Alkhazov:1997zz,kiselev05} of $2.33(10)$.

Table~\ref{p_radius} presents analogous results for the point-proton radius.
Convergence behavior and comparisons with the NCSM are also analogous.
Even though the
protons belong to the core and not to the halo, the extension of the halo plays
an important role for the point-proton radius.  It displaces the center  
of mass of the core from the center of mass of the whole system, increasing the 
point-proton radius as it is easily seen in Eq.~\eqref{eq:rpp}. 
Our most complete results of $1.90(2)$ fm  is on the lower side but compatible with 
the bounds for the point-proton radius [$1.938(23)$ fm] as evaluated in Ref.~\cite{Bacca:2012up}.
 
It is important to point out that while the use of the $\lambda_{\rm SRG}$=1.5 fm$^{-1}$ 
SRG parameter produces a softer $NN$ potential and hence faster convergence, 
it is known that at this resolution scale there are 
significant SRG-induced $3N$ forces as well as SRG-induced two- and three-body contributions to the radii. 
Within the present calculations we are
disregarding such induced terms. Therefore, the results obtained with this resolution scale are expected to be 
far from realistic and they should be understood as an instrument to study the NCSMC approach
rather than as realistic predictions
for the $^6$He nucleus.  
 
\begin{figure}[t]
      \includegraphics[width=6.5cm,clip=,draft=false,angle=-90]{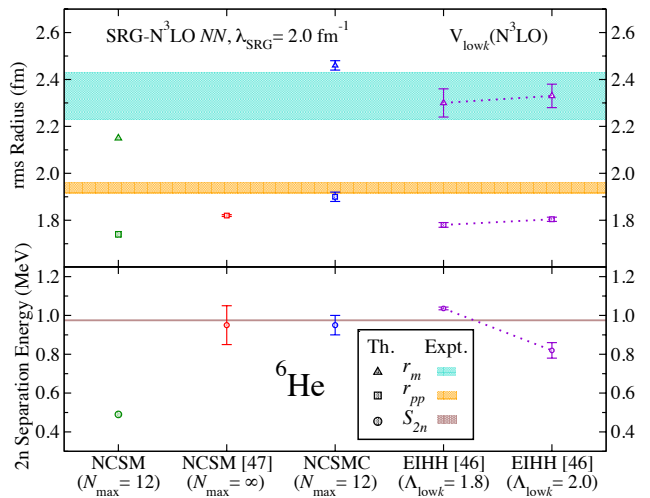}  
\caption{ (Color online) NCSM (green symbols) and NCSMC (blue symbols) rms matter 
(triangles) and point-proton (squares) radii, and two-neutron separation energy
(circles), obtained using the SRG-N$^3$LO $NN$ interaction with $\lambda_{\rm SRG}=2.0$ fm$^{-1}$
in the largest HO model space.
Also shown are the 
infinite-basis extrapolations 
from Ref.~\cite{Saaf2014} (red symbols) %. 
and the 
EIHH results   
from Ref.~\cite{Bacca:2012up} (indigo symbols)
at 
the resolution scales $\Lambda_{{\rm low}k}=1.8$, and 2.0 fm$^{-1}$.
The range of experimental values 
are represented by horizontal bands (see text for more details).
}
\label{fig:conv}
\end{figure}

A summary of the rms radii obtained for the more realistic 
$\lambda_{\rm SRG}$=2.0 fm$^{-1}$ interaction is presented in Table~\ref{summary} and visualized in Fig.~\ref{fig:conv} 
together with the corresponding results for the separation energy,
the infinite-basis extrapolations from Ref.~\cite{Saaf2014}, and the  
effective interaction hyperspherical harmonics (EIHH) calculations from Ref.~\cite{Bacca:2012up}, based on the $V_{{\rm low}k}$(N$^3$LO) $NN$ interaction at 
the resolution scales $\Lambda_{{\rm low}k}=1.8$, and 2.0 fm$^{-1}$.
(The results presented Table~\ref{summary} have been obtained with improved accuracy and supersede those shown in 
Table II of Ref.~\cite{rom2016}, where the labeling of the HO model space size was also incorrectly reported to be lower by two units.) 

\begin{table}[t]
\caption{
Summary of the results presented in Fig.~\ref{fig:conv}, with $\Lambda_{{\rm low}k}$ in units of fm$^{-1}$. 
See text 
for further details.  
}
\begin{ruledtabular}
\begin{tabular}{ l l c c c}
           & &$S_{2n}$ (MeV) & $r_m$ (fm) & $r_{pp}$ (fm)\\ 
\hline
NCSM &($N_{\rm max}=12$)    &  0.49   & 2.15 & 1.74 \\
NCSM~\cite{Saaf2014} &($N_{\rm max}=\infty$) &   0.95(10)     &   --       &1.820(4)\\
NCSMC &($N_{\rm max}=12$)   &  0.94(5)   & 2.46(2) & 1.90(2) \\
EIHH~\cite{Bacca:2012up} &($\Lambda_{{\rm low}k}=1.8$) &1.036(7) & 2.30(6) &1.78(1) \\
EIHH~\cite{Bacca:2012up} &($\Lambda_{{\rm low}k}=2.0$) &0.82(4) & 2.33(5) &1.804(9) \\
\hline
Expt.    &&   0.975   & 2.33(10) & 1.938(23)\\
\end{tabular}
\end{ruledtabular}
\label{summary}
\end{table}
The two-nucleon separation energy obtained within the NCSMC is close to its empirical value, 
and the computed $r_m$ %matter 
and $r_{pp}$
%point-proton 
radii are, respectively, at the upper end of and on the lower side but compatible with  their experimental bands. 
Interestingly, our point-proton radius  is substantially larger than both the extrapolated value of S\"a\"af et al., which ``calls for further investigations"~\cite{Saaf2014}, and the  
EIHH result of Bacca et al.~\cite{Bacca:2012up}.  
This latter calculation also yields a matter radius smaller than ours though within the experimental bounds. 
The present combination of $S_{2n}$ and $r_{pp}$ values are more in line with the Green's function Monte Carlo results of Ref.~\cite{Pieper2008}, based on $NN$+$3N$ forces constrained to reproduce the properties of light nuclei including $^6$He.

\subsection{$^4$He+$n$+$n$ continuum}
We investigated the low-lying $\alpha$+$n$+$n$ continuum for partial waves up to $J^{\pi}$ =  2$^{\pm}$ 
by solving the set of Eqs.~\eqref{eq:system} with the boundary conditions
from Eq.~\eqref{eq:ext_cont}. The eigenphase shifts were extracted from the 
diagonalization of the three-body scattering matrix $S_{\nu K,\nu' K'}$. 

Convergence of the results with respect to the HO model-space size and the parameters  
used to perform the matching between the solutions in the internal region and the asymptotic wave functions within the $R$-matrix approach 
was reached at similar values as those used in our previous study of Ref.~\cite{Romero-Redondo:2014fya}, lacking the contribution from square-integrable eigenstates of the composite system. 
Specifically, our best results were obtained at $N_{\mbox{max}}$=12, which is the maximum computationally accessible HO model space size, and  
interested readers can find a complete list of the remaining parameters  for each channel in Appendix~\ref{ap:parameters}.
\begin{figure*}[t]
\begin{center}
  \includegraphics[width=0.38\textwidth,angle=-90]{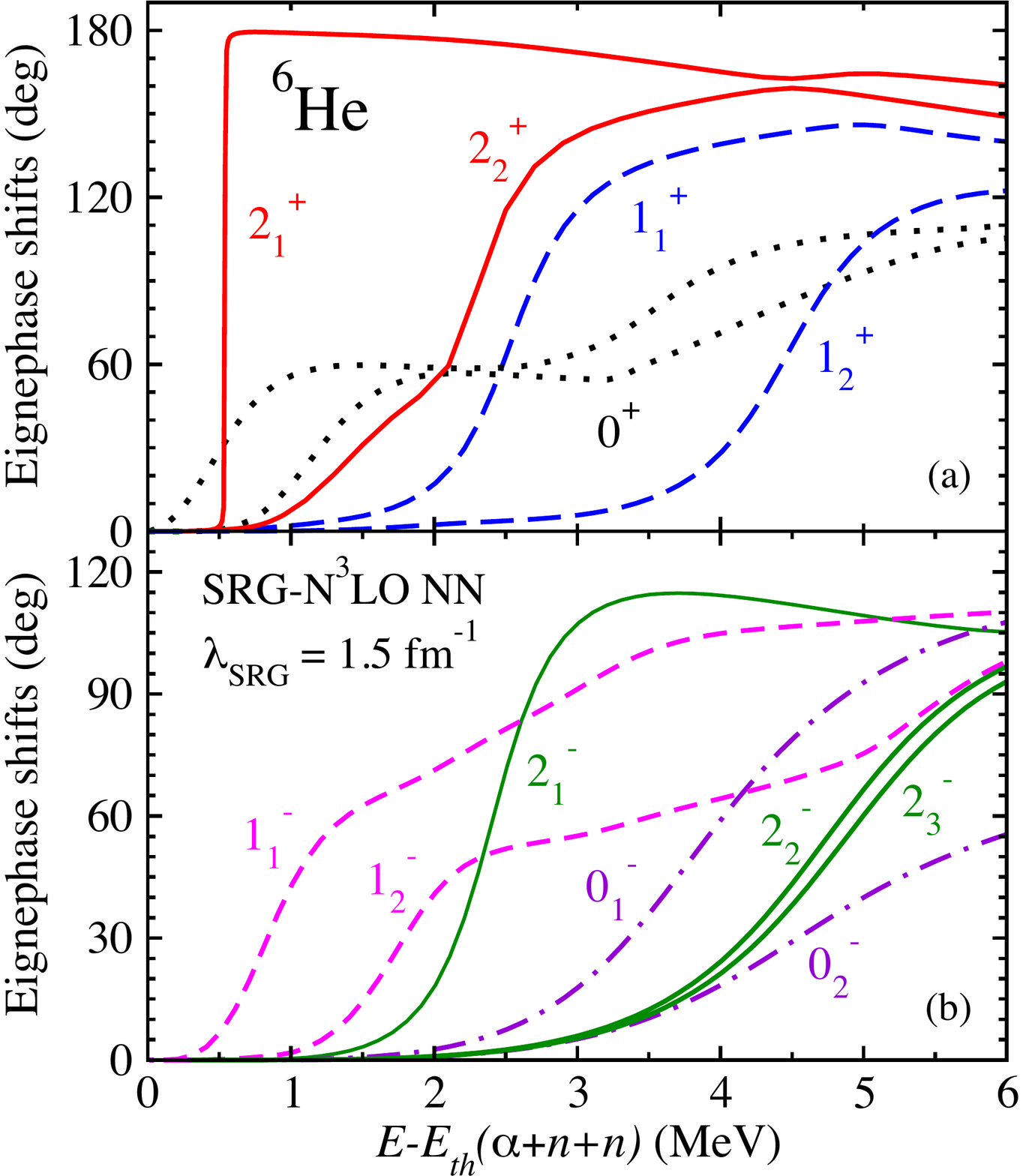}
  \includegraphics[width=0.38\textwidth,angle=-90]{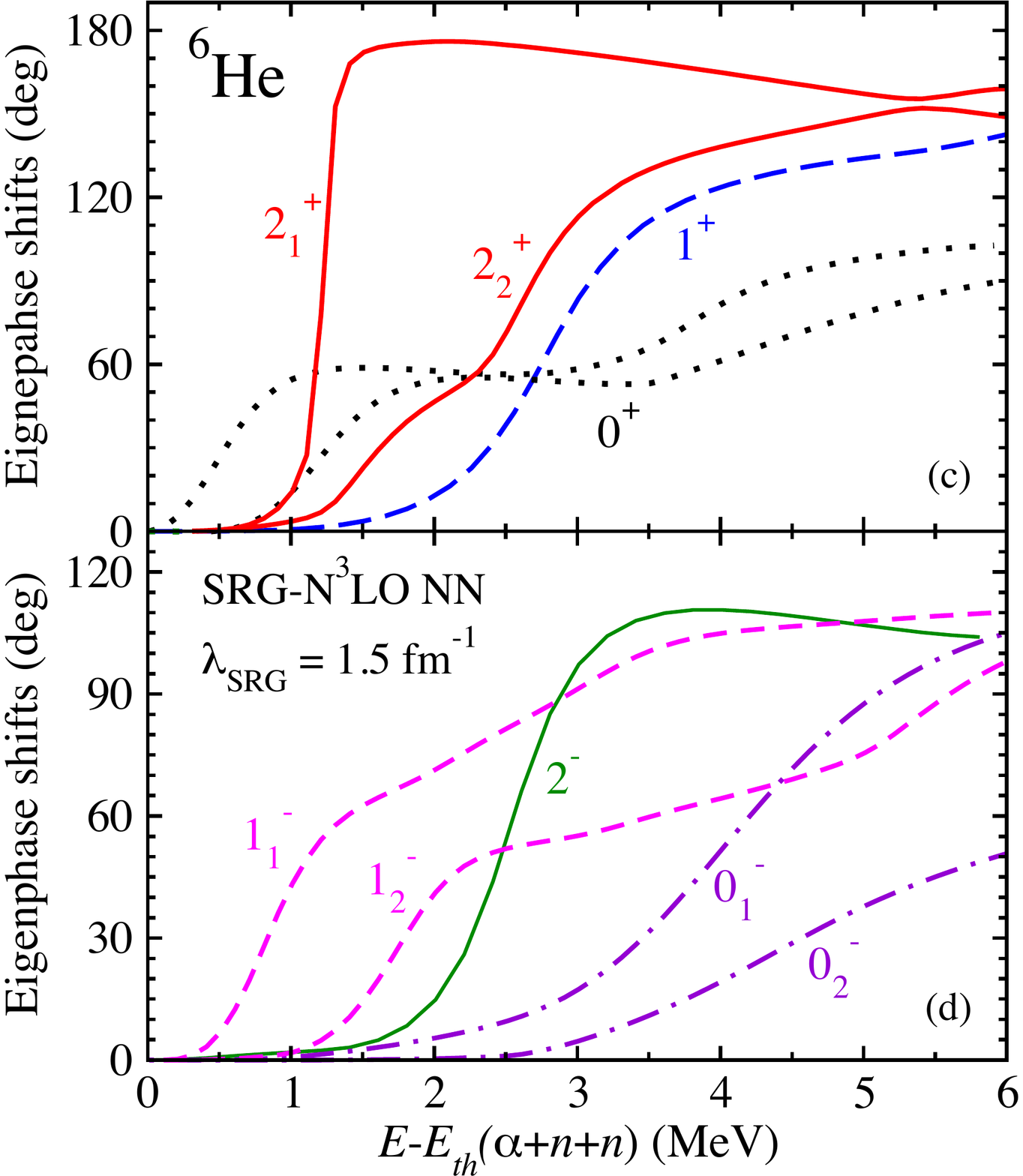}
\end{center}
\caption{Attractive eigenphase shifts below 6 MeV above the two-neutron emission threshold [$E_{th}$($\alpha$+$n$+$n$)]  computed within the NCSMC [panels (a) and (b)]
and within the more limited model space spanned by the $^4$He(g.s.)+$n$+$n$ cluster basis alone of Ref.~\cite{Romero-Redondo:2014fya} [panels (c) and (d)], using the SRG-evolved N$^3$LO $NN$ potential with $\lambda_{\rm SRG}=$ 1.5 fm$^{-1}$.
 We show positive parity states in panels (a) and (c), and negative
parity states in panels (b) and (d).}
\label{summary_eigenphase_shift}
\end{figure*}
\begin{figure}[b]
      \includegraphics[width=6.cm,clip=,draft=false,angle=-90]{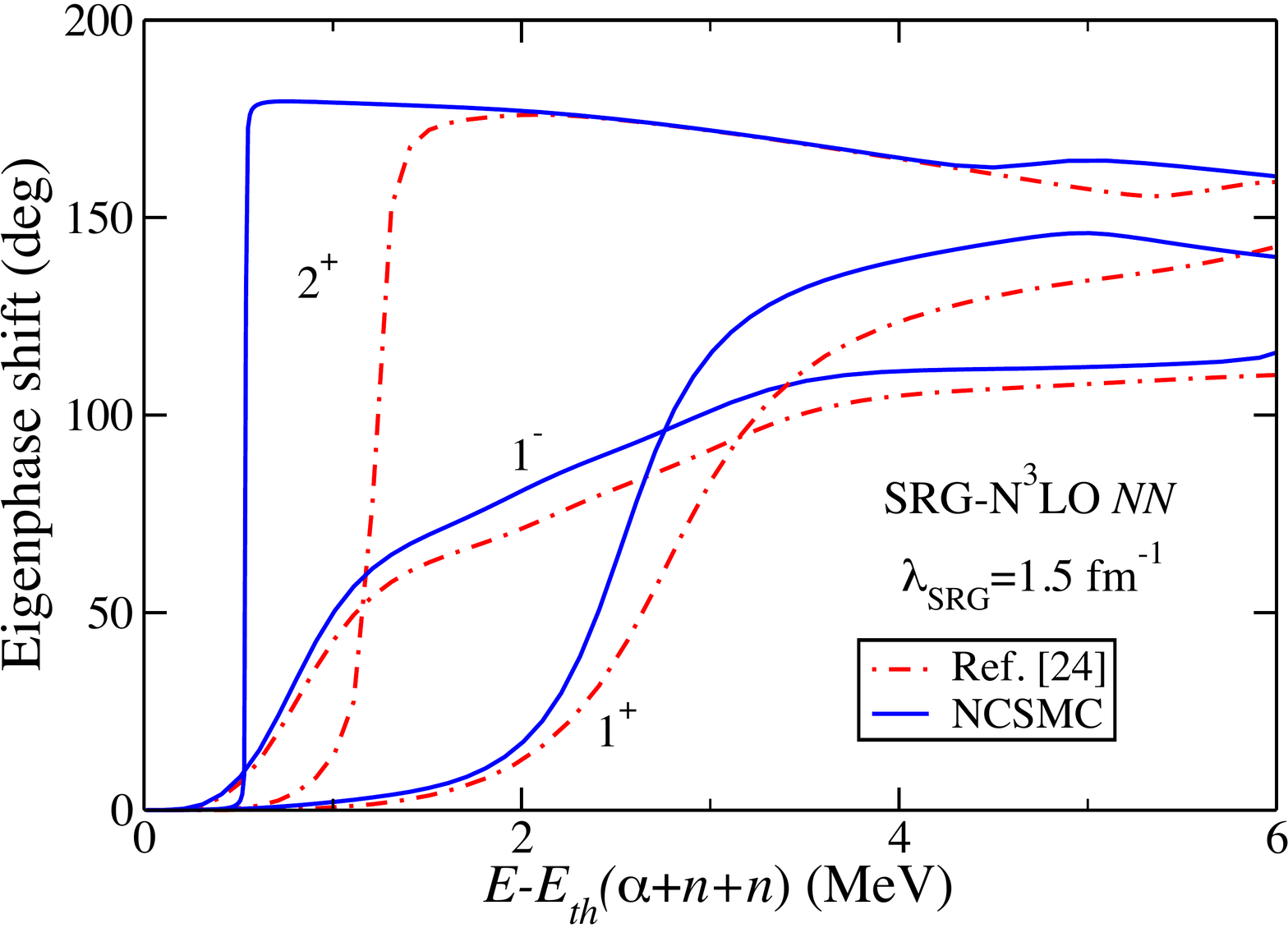}  
\caption{Eigenphase shifts for the $J^\pi=1^{\pm}$ and 2$^{+}$ channels
computed within the NCSMC (blue solid line) and within the more limited model space spanned by the $^4$He(g.s.)+$n$+$n$ cluster basis alone of Ref.~\cite{Romero-Redondo:2014fya} (red dot-dashed line) using the SRG-evolved N$^3$LO $NN$ potential with $\lambda_{\rm SRG}=$ 1.5 fm$^{-1}$. 
}
\label{eigenphase_shift}
\end{figure}
\begin{figure}[b]
      \includegraphics[width=6.cm,clip=,draft=false,angle=-90]{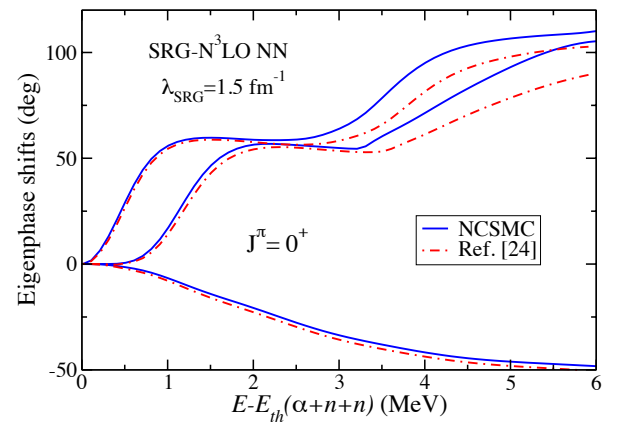}  
\caption{Eigenphase shifts for the $J^\pi$= 0$^+$ channel 
computed within the NCSMC (blue solid line) and within the more limited model space spanned by the $^4$He(g.s.)+$n$+$n$ cluster basis alone of Ref.~\cite{Romero-Redondo:2014fya} (red dot-dashed line) using the SRG-evolved N$^3$LO $NN$ potential with $\lambda_{\rm SRG}=$ 1.5 fm$^{-1}$. 
}
\label{eigenphase_shift2}
\end{figure}
     
In Fig.~\ref{summary_eigenphase_shift}(a) and (b), we present a summary of the 
most relevant attractive eigenphase shifts below $6$ MeV obtained for
the $\lambda_{\rm SRG}=$ 1.5 fm$^{-1}$ interaction within the NCSMC  by including the first nine positive-parity and six negative-parity $J\le2$ square-integrable eigenstates of the composite system.  
This figure can be compared with Fig.~1 of 
Ref.~\cite{Romero-Redondo:2014fya} -- for convenience shown again in Fig.~\ref{summary_eigenphase_shift}(c) and (d) -- which presents analogous results   
computed within the more limited model space spanned by the $^4$He(g.s.)+$n$+$n$
cluster basis alone. 
Although the
qualitative behavior of the eigenphase shifts is similar, within the NCSMC
the centroid values of all resonances tend to be shifted to lower energies and  the resonance widths tend to shrink due to the effect of the inclusion of
discrete eigenstates of the composite system. 
The most significant change is observed for the first 2$^+$ resonance, which becomes much sharper (with a width of $\Gamma = 15$ keV) and is shifted to lower energies (with the new centroid at 0.536 MeV). This behavior suggests a likely significant influence of the chiral 3N force on this state. The effect in other partial waves is more modest. In particular, the $1^-$ eigenphase shift does not change significantly, excluding core-polarization effects as the possible origin of a low-lying soft dipole mode. This can more readily be  observed in Fig.~\ref{eigenphase_shift} and \ref{eigenphase_shift2}, where we show a direct comparison between the present results and those of Ref.~\cite{Romero-Redondo:2014fya}  for the lowest resonances in the 1$^{\pm}$ and 2$^+$ channels and
for the lowest three eigenphase shifts in the 0$^+$ channel, respectively.   
The repulsive eigenphase shift in the 0$^+$ channel corresponds to the ground state of $^6$He,
and the small difference between the calculations is related to the difference in the binding energy, 
as it was shown in Table~\ref{energy}.

\begin{figure}[t]
\includegraphics[width=6cm,clip=,draft=false,angle=-90]{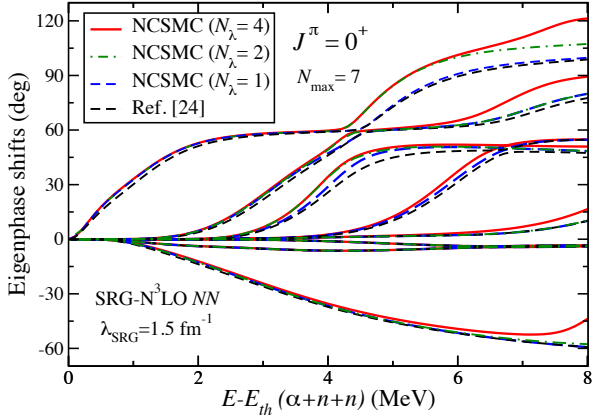}
\caption{
Convergence with respect to the number ($N_\lambda$) of square-integrable eigenstates of the composite system included in the NCSMC calculation of the eigenphase shifts for the 0$^+$ channel,  using the SRG-evolved N$^3$LO $NN$ potential with $\lambda_{\rm SRG}=$ 1.5 fm$^{-1}$. Also shown are the results from Ref.~\cite{Romero-Redondo:2014fya}, corresponding
to the inclusion of zero composite states (cluster basis alone).   
}
\label{convergence_NCSMC} 
\end{figure}
The convergence of the eigenphase shifts with respect to the number of eigenstates of the
composite system included in the calculation was found to be very fast.  
The mere inclusion of the lowest eigenstate is in general sufficient to obtain 
reasonable convergence in the low-energy region. 
As an example, we show in Fig.~\ref{convergence_NCSMC} the convergence pattern
of the $J^{\pi}=0^+$ eigenphase shifts with respect to the number of NCSM eigenstates of the composite system
for a small model space of size $N_{\rm max}$= 7. %
Two eigenstates are already sufficient for obtaining convergence up to 5 MeV. For energies 
below 3 MeV, a single eigenstate is enough. This convergence behavior is of course related to the value of the eigenenergies associated with the included square-integrable eigenstates.  The further the eigenvalue is from the energy under consideration, the smaller the contribution to the eigenphase shifts from the corresponding eigenstate. (The eigenenergies of all positive- and negative-parity eigenstates included in the $N_{\rm max}=12$ calculations are shown in tables~\ref{pos_par} and \ref{neg_par}, respectively.) For comparison, the 
eigenphase shifts of Ref.~\cite{Romero-Redondo:2014fya}, calculated within the cluster basis alone, are also shown (corresponding to zero eigenstates included).   
\begin{figure}[b]
\includegraphics[width=7cm,clip=,draft=false,angle=90]{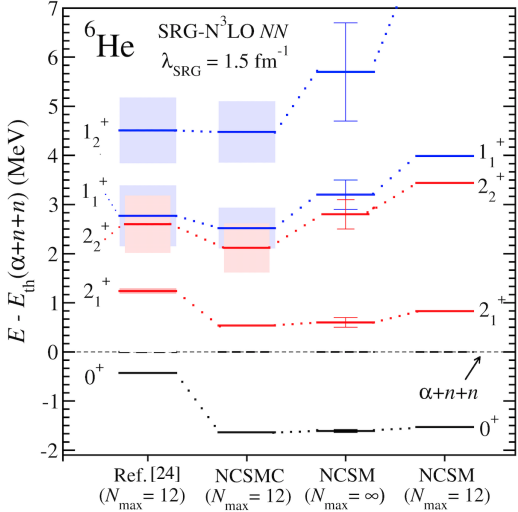}
\caption{Spectrum of low-lying energy levels of the $^6$He nucleus for 
$\lambda_{\rm SRG}=1.5$  
as obtained within the $^4$He(g.s.)+$n$+$n$ cluster basis of Ref.~\cite{Romero-Redondo:2014fya}, the NCSMC, and within the NCSM by treating the $^6$He excited states as bound states. For the NCSM , we show both
the energy levels at $N_{\rm max}$=12 and the results of an extrapolation to 
the infinite model space, performed through the exponential fit function from Eq.~\eqref{expo_fit}.  
}
\label{spectrum_methods} 
\end{figure}

From the calculated eigenphase shifts, it is possible to extract information about the resonances by      
calculating the centroids $E_R$ and widths $\Gamma$
as the values of $E_{\rm kin}=E-E_{th(\alpha+n+n)}$ for which
  the first derivative $\delta^\prime(E_{kin})$ of the eigenphase shifts is
  maximal and $\Gamma{=}2/\delta^\prime(E_R)$, respectively \cite{Thompson2009}.
The resulting low-lying $^6$He spectrum of energy levels for the SRG-evolved N$^3$LO $NN$ interaction with $\lambda_{\rm SRG}=1.5$ fm$^{-1}$    
  is shown in Fig.~\ref{spectrum_methods}. There, we compare
 the present NCSMC results with the spectra computed within the cluster basis
alone~\cite{Romero-Redondo:2014fya}, and  within the NCSM (i.e., by treating the $^6$He excited states as bound states).  
Besides the results at  $N_{\rm max}=12$, for the NCSM we also show the spectrum obtained by extrapolation to the infinite HO model space  using the exponential form of Eq.~\eqref{expo_fit}. 
Note that, while for the results of Ref.~\cite{Romero-Redondo:2014fya} and the NCSMC the
resonances are represented by their centroids (solid line) and width (shaded area), for 
the NCSM we only show the energy levels and associate estimated uncertainty of the extrapolation.      
Indeed, such a bound-state technique does not yield resonance widths. While broad, higher-energy states 
 such as the 1$^+_2$ resonance are well described already within a $^4$He(g.s)+$n$+$n$ picture and very narrow resonances such as the first 2$^+$ can already be explained within the bound-state approximations of the NCSM, for other intermediate levels both short-range many-body correlations and continuum degrees of freedom play an important role.
\begin{figure}[t]
      \includegraphics[width=8cm,clip=,draft=false,angle=0]{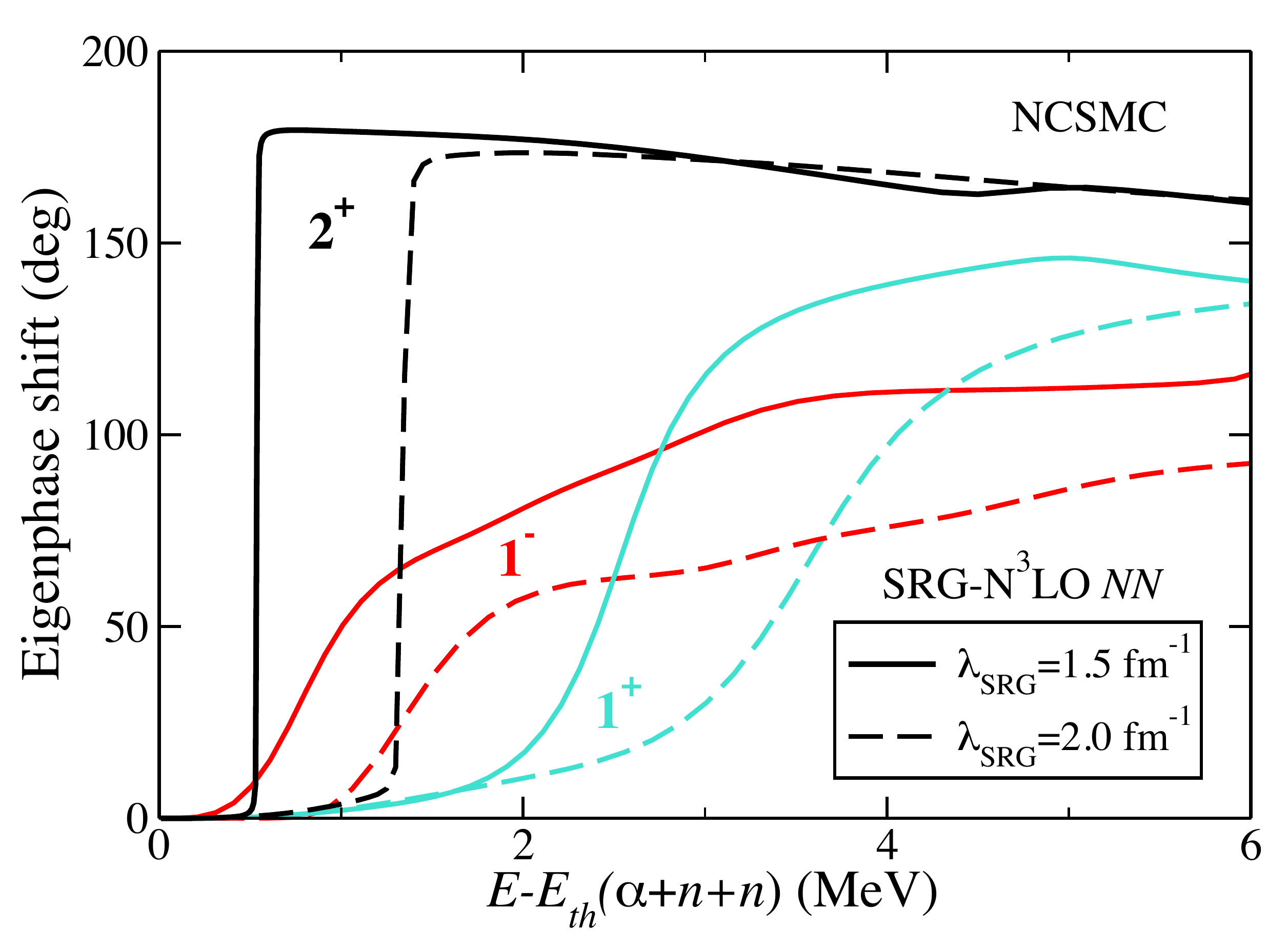}  
\caption{Eigenphase shifts for the$J^\pi=1^{\pm}$ and 2$^{+}$ channels
computed within the NCSMC using the SRG-evolved N$^3$LO $NN$ potential with $\lambda_{\rm SRG}=$ 1.5 fm$^{-1}$ (solid lines) and
  $\lambda_{\rm SRG}=$ 2.0 fm$^{-1}$ (dashed lines).  
}
\label{eigenphase_shift3}
\end{figure}
\begin{figure}[t]
      \includegraphics[width=7cm,clip=,draft=false]{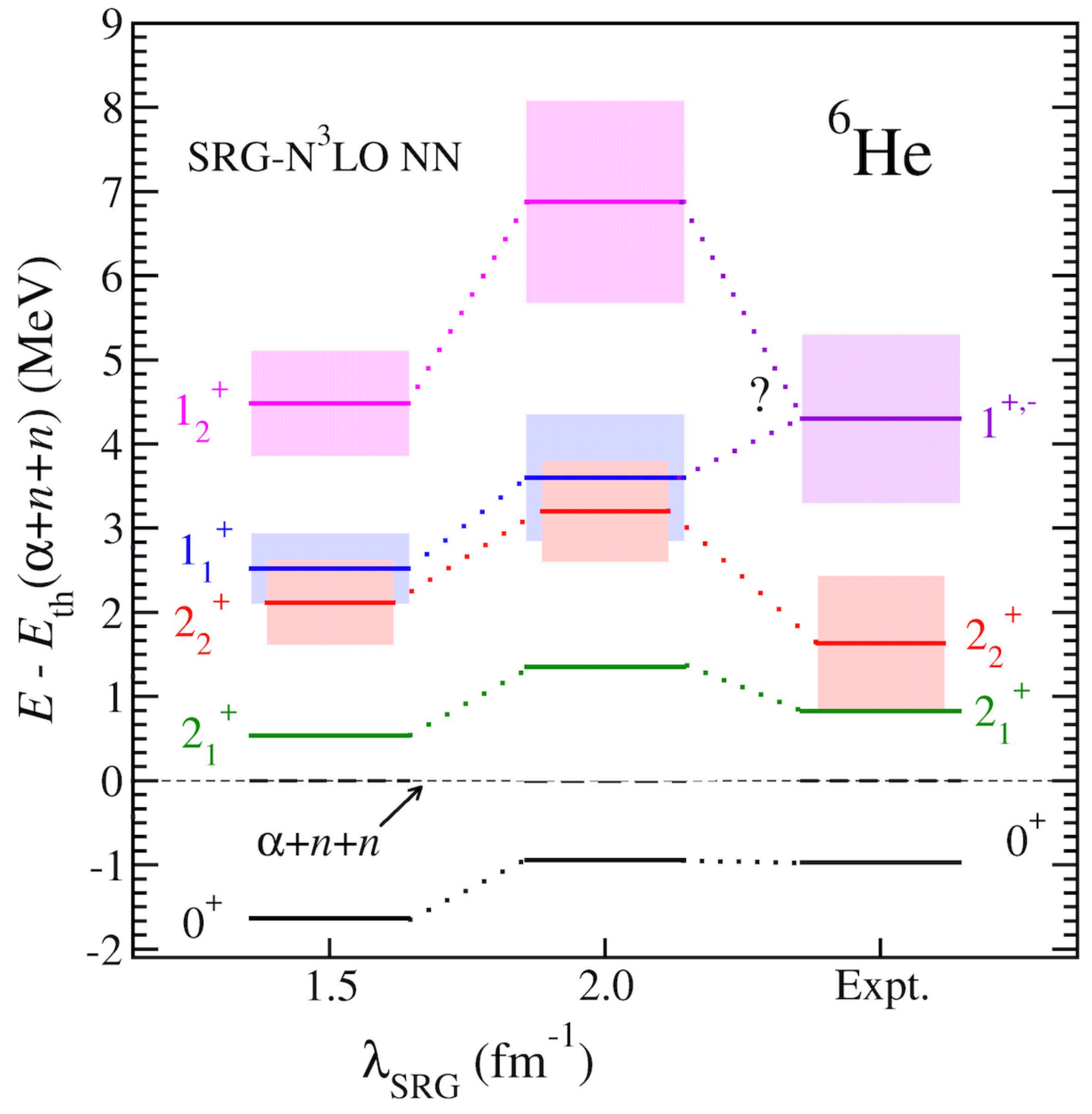}   
\caption{Spectra of low-lying energy levels of the $^6$He nucleus computed within the NCSMC using the SRG-evolved N$^3$LO $NN$ potential with $\lambda_{\rm SRG}=$ 1.5 fm$^{-1}$  and $\lambda_{\rm SRG}=$ 2.0 fm$^{-1}$ compared to the most recent experimental spectrum of Ref.~\cite{Mougeot:2012aq}. 
}
\label{spectrum_lambda}
\end{figure}

 The harder $NN$ interaction obtained with the SRG resolution scale of $\lambda=2.0$ fm$^{-1}$ produces a qualitatively similar picture, but with higher-lying and wider resonances. This is highlighted in 
Figs.~\ref{eigenphase_shift3} and \ref{spectrum_lambda}, showing respectively 
the eigenphase shifts for  the $J^\pi=1^{\pm}$ and 2$^+$ channels, and  a comparison of the computed energy levels with 
the most recent experimental spectrum of Ref.~\cite{Mougeot:2012aq}.
The observed dependence on the value of the SRG resolution scale provides an estimate of the effect of induced $3N$ (and higher order) 
forces, which  have been disregarded in the present study and are crucial to restore the formal unitarity of the adopted SRG transformation of the Hamiltonian. 
More in general, the inclusion of 3N forces (including the initial chiral $3N$ force) is indispensable to arrive at an accurate description of the spectrum as a whole. 
Indeed, while the SRG-evolved $NN$ interaction with $\lambda=2.0$ fm$^{-1}$ provides a realistic description of the energy and structure of the $^6$He ground state, neither of the two adopted resolution scales describes accurately the spectrum of the low-energy excited states.  At the same time, based on these results we conjecture that the parity of the $J=1$ resonance  populated at SPIRAL through the $^8$He($p,^3$He)$^6$He* two-neutron transfer reaction~\cite{Mougeot:2012aq} is likely positive, making it less probable that this state is the soft-dipole mode called for by Refs.~\cite{PhysRevLett.85.262}
and \cite{Nakamura2000209}.

\section{Conclusions} 
\label{sec:conclusions}
We presented the extension of the {\it ab initio} no-core shell model with
continuum to the treatment of bound and continuum nuclear systems in the  
proximity of a three-body breakup threshold. This approach takes
simultaneously into account  
both many-body short-range correlations and clustering degrees of freedom, allowing for a
comprehensive  {\it ab initio} description of nuclear systems presenting a three-cluster 
configuration such as Borromean halo nuclei and light-nuclei reactions with three nuclear fragments in either entrance or exit channels.

After introducing the NCSMC ansatz for systems characterized by a three-cluster asymptotic behavior, we discussed the dynamical equations, and gave the algebraic expressions of the overlap and Hamiltonian couplings between the discrete and continuous NCSMC basis states for the particular case of core+$n$+$n$ systems.   Further, we discussed the procedure adopted for the solution of the three-cluster dynamical equations for bound and scattering states, and explained how we  calculate the probability density, and the matter and point-proton root-mean-square radii starting from the obtained NCSMC solutions for core+$n$+$n$ systems. 
The new formalism was then applied to conduct a comprehensive study of many-body correlations and $\alpha$-clustering in the ground-state and low-lying energy continuum of the Borromean $^6$He nucleus using the chiral N$^3$LO $NN$ potential or
Ref.~\cite{N3LO} softened via the similarity renormalization group
method~\cite{PhysRevC.75.061001,
PhysRevC.77.064003,Wegner1994}. 

Calculations were carried out using a soft ($\lambda_{\rm SRG}=1.5$ fm$^{-1}$) SRG resolution scale to allow for a direct comparison with the 
results obtained in the more limited studies of Refs.~\cite{Quaglioni2013,Romero-Redondo:2014fya}, based solely on the
three-cluster portion of the NCSMC basis. 
While working within the $^4$He(g.s.)+$n$+$n$ microscopic cluster basis 
it is possible to reproduce the correct asymptotic behavior of the $^6$He wave function,
we demonstrated that additional short-range six-body correlations (included in the form of square-integrable eigenstates of the composite $^6$He system) are necessary to 
correctly describe also the interior of the wave function for both the ground and scattering states. In particular, a significant portion of the ground-sate energy and the narrow width of the first 2$^+$ resonance stem from many-body correlations that, in a microscopic-cluster picture, can be interpreted as core-excitation effects.

A second and physically more interesting potential ($\lambda_{\rm SRG}=2.0$ fm$^{-1}$) 
was also used. Though the inclusion of $3N$ forces (currently underway) remains crucial to restore the formal unitarity of the adopted SRG transformation of the Hamiltonian and arrive at an accurate description of the spectrum as a whole, the present results demonstrated that rms matter and point-proton radii compatible with experiment can be obtained starting from a soft $NN$ interaction reprodu cing the $^6$He small binding energy.  

In the future we plan to reexamine the {\em ab initio} calculation of the $^6$He $\beta$-decay half-life, first carried out in Ref.~\cite{PhysRevC.65.054302}, in the context of chiral effective field theory using wave functions with proper asymptotic behavior. This work also sets the stage for the {\em ab initio} study of the $^4$He$(2n,\gamma)^6$He radiative capture and is a stepping stone in the calculation of the $^3$H$(^3{\rm H},2n)^4$He fusion.

%\newpage

\appendix

\begin{widetext}
\section{Norm and Hamiltonian kernels}
\label{ap:kernels}

Here we present the explicit expressions for the NCSMC 
Hamiltonian and norm kernels entering Eqs.~\eqref{eq:orthoH} and \eqref{eq:orthowf}.
There, the square and inverse-square root of the NCSMC norm kernel, ${\bold N}^{\pm\frac{1}{2}}$, can be written as   
\begin{align}\label{eq:Norm}
  ({\bold N}^{\pm\frac{1}{2}})^{\lambda\lambda^\prime}_{\nu xy,\nu^\prime x^\prime y^\prime} &= 
\left(
\begin{array}{ccc}
0  && 0 \\[2mm]
0  && \delta_{\nu\nu'}\frac{\delta(x-x')}{xx'}\frac{\delta(y-y')}{yy'}
- \delta_{\nu\nu'} \delta_{n^\prime_x n_x} \delta_{n^\prime_y n_y}R_{n_x\ell_x}(x) R_{n_x\ell_x}(x')R_{n_y\ell_y}(y) 
R_{n_y\ell_y}(y')
\\[2mm]
		\end{array}
  \right)
\nonumber\\[3mm]
&+\left(
\begin{array}{ccc} 
\delta_{\lambda\tilde{\lambda}} && 0\\[2mm]
0 &&  R_{n_x\ell_x}(x)R_{n_y\ell_y}(y)\delta_{\nu\tilde{\nu}}
\end{array}
\right)
( {\bold N}^{\pm\frac{1}{2}})^{\tilde{\lambda}\tilde{\lambda}'}_{\tilde{\nu}n_x n_y,\tilde{\nu}'n_x'n_y'}
\left(
\begin{array}{ccc} 
\delta_{\tilde{\lambda}'\lambda'} && 0\\[2mm]
0 &&  R_{n_x'\ell_x'}(x')R_{n_y'\ell_y'}(y')\delta_{\tilde{\nu}'\nu'}
\end{array}
\right),
\end{align}
where the sum over the repeating indexes $\tilde\lambda, \tilde\nu,
n_x, n_y, \tilde\lambda^\prime, \tilde\nu^\prime,
n_x^\prime$, and $n_y^\prime$ is implied, and the notation
\begin{align} 
({\bold N}^{\pm\frac12})^{\lambda\lambda^\prime}_{\nu n_x
  n_y,\nu^\prime n_x^\prime n_y^\prime}
=\left(
\begin{array}{ccc}  
({\bold N}^{\pm\frac12})^{(11)}_{\lambda\lambda'}&&({\bold
  N}^{\pm\frac12})^{(12)}_{\lambda\nu' n_x' n_y'}\\[2mm]
({\bold N}^{\pm\frac12})^{(21)}_{\lambda'\nu n_x n_y} && ({\bold
  N}^{\pm\frac12})^{(22)}_{\nu n_x n_y, \nu^\prime n_x^\prime n_y^\prime}  
\end{array}
\right)
\label{eq:NPMmodel}
\end{align}
stands for the matrix elements of the square and inverse-square root
of the NCSMC norm kernel within the model space, which are computed
from the NCSMC model-space norm kernel
\begin{align} 
{\bold N}^{\lambda\lambda'}_{\nu n_x n_y,\nu' n_x' n_y'}
=\left(
\begin{array}{ccc}  
\delta_{\lambda\lambda'}&&\bar g_{\lambda \nu' n_x' n_y'}\\[2mm]
\bar g_{\lambda' \nu n_x n_y} && \delta_{\nu\nu'}\delta_{n_x
  n_x'}\delta_{n_y n_y'}  
\end{array}
\right)
\label{eq:Nmodel}
\end{align}
using the spectral theorem.
The orthogonalized Hamiltonian within the model space can then be calculated as follows    
\begin{equation} 
\overline{\bold H}^{\lambda\lambda'}_{\nu n_x n_y,\nu' n_x' n_y'}=
\left(
\begin{array}{ccc}
\overline{\bold H}^{(11)}_{\lambda\lambda'} && \overline{\bold H}^{(12)}_{\lambda\nu'n_x'n_y'}\\ 
\overline{\bold H}^{(21)}_{\lambda'\nu n_x n_y}&&\overline{\bold H}^{(22)}_{\nu n_x n_y, \nu'n_x'n_y'} 
\end{array}
\right) 
= ( {\bold N}^{-\frac{1}{2}})^{\lambda\tilde{\lambda}}_{\nu n_x
  n_y,\tilde{\nu}\tilde n_x \tilde n_y}{\bold
  H}^{\tilde{\lambda}\tilde{\lambda}'}_{\tilde{\nu}\tilde n_x \tilde
  n_y,\tilde{\nu}' \tilde n_x' \tilde n_y'} 
({\bold N}^{-\frac 12})^{\tilde{\lambda}'\lambda'}_{\tilde{\nu}'\tilde n_x' \tilde
  n_y',\nu' n_x' n_y'},
\label{eq:ortgHms}
\end{equation}
where the sum over the repeating indexes $\tilde\lambda, \tilde\nu,
\tilde n_x, \tilde n_y, \tilde\lambda^\prime, \tilde\nu^\prime,
\tilde n_x^\prime$, and $\tilde n_y^\prime$ is, once again, implied,
and 
\begin{align}
{\bold H}^{\lambda\lambda^\prime}_{\nu n_x n_y, \nu^\prime n_x^\prime n_y^\prime} = 
	\left(
		\begin{array}{lcl}
      	  E_\lambda\delta_{\lambda\lambda^\prime} & &
          \bar{h}_{\lambda\nu^\prime n_x' n_y'} \\ [2mm] 
		   \bar{h}_{\lambda^\prime\nu n_x n_y}
                   &&\overline{\mathcal{H}}_{\nu n_x n_y, \nu^\prime
                     n_x\ n_y'} 
		\end{array}
	\right)\,,
	\label{eq:Hms}
\end{align}
is the model-space component of the NCSMC Hamiltonian kernel of
Eq.~\eqref{eq:H}. We note that the coupling form factors in
configuration space, $\bar{h}_{\lambda\nu}(x,y) = [h{\mathcal
  N}^{-\frac12}]_{\lambda\nu}(x,y)$ are related to those in the model
space, $\bar{h}_{\lambda\nu n_x n_y}$, through Eqs.~\eqref{eq:h} and
\eqref{eq:transf}, and the lower-diagonal block is the model-space
component of the orthonormalized integration kernel of Eq.~\eqref{eq:formalism_90}.
Additional details on how this kernel is computed can be found in Ref.~\cite{Quaglioni2013}, 
where we introduced the formalism for the description of three-cluster dynamics based solely on expansions over three-cluster channels states of the type of Eq.~\eqref{eq:3bchannel}. 

Finally, in the following we provide detailed expressions for the
blocks forming the orthogonalized NCSMC Hamiltonian of
Eq.~\eqref{eq:orthoH}, including the terms that extend beyond the
HO model space $P$. 
In particular, in the following we will use the notation
$n\in P$ to indicate that the radial quantum number $n \le
N_{\rm max}$. 
Note that for the upper diagonal bock there are not additional terms that reach beyond the 
the HO model space and, therefore, it is trivially given by the upper diagonal block of 
Eq.~\eqref{eq:ortgHms}.   
\begin{align}
&\overline{\bold H}^{(12)}_{\lambda\nu^\prime}( x^\prime, y^\prime)
=\sum_{n_x'n_y'}R_{n_x'\ell_x'}(x')  
R_{n_y'\ell_y'}(y')\overline{\bold H}^{(12)}_{\lambda\nu'n_x'n_y'}\nonumber\\
&\qquad+\sum_{\tilde{\lambda}}({\bold N}^{-\frac 12})^{(11)}_{\lambda\tilde{\lambda}}
\left[\sum_{n_y'\in P} R_{N+1\ell_x'}(x')g_{\tilde{\lambda}\nu'N n_y'} 
T^{\ell_x'}_{N N+1}R_{n_y'\ell_y'}(y')+ 
\sum_{n_x'\in P} R_{N+1\ell_y'}(y')g_{\tilde{\lambda}\nu'n_x'N}
T^{\ell_y'}_{N N+1}R_{n_x'\ell_x'}(x') \right] \nonumber\\    
&\qquad+\sum_{\tilde{\nu}}\sum_{\tilde n_x\tilde n_y n_y'\in P}({\bold
  N}^{-\frac 12})^{(12)}_{\lambda\tilde{\nu}\tilde n_x\tilde n_y } 
\left(\frac{1}{2}\Lambda^{-\frac 12}_{\tilde{\nu}\tilde n_x\tilde n_y,\nu'N n_y'} 
+\frac{1}{2}\Lambda^{\frac 12}_{\tilde{\nu}\tilde n_x\tilde n_y,\nu'N n_y'} 
\right)
T^{\ell_x'}_{N N+1}R_{N+1\ell_x'}(x')R_{n_y'\ell_y'}(y')\nonumber\\ 
&\qquad+\sum_{\tilde{\nu}}\sum_{\tilde n_x\tilde n_y n_x'\in P}({\bold
  N}^{-\frac 12})^{(12)}_{\lambda\tilde{\nu}\tilde n_x\tilde n_y}
\left(\frac{1}{2}\Lambda^{-\frac 12}_{\tilde{\nu}\tilde n_x\tilde n_y, \nu'n_x'N} 
+\frac{1}{2}\Lambda^{\frac 12}_{\tilde{\nu}\tilde n_x\tilde n_y, \nu'n_x'N} 
\right)
T^{\ell_y'}_{N N+1}R_{N+1\ell_y'}(y')R_{n_x'\ell_x'}(x') \,,
%\nonumber\\[5pt]  
\end{align}
\begin{align}
&\overline{\bold H}^{(21)}_{\lambda^\prime\nu}(x,y) 
=\sum_{n_xn_y}R_{n_x\ell_x}(x)  
R_{n_y\ell_y}(y)\overline{\bold H}^{(21)}_{\lambda'\nu n_xn_y}\nonumber\\
&\qquad+\sum_{\tilde{\lambda}}\left[\sum_{n_y\in P} R_{N+1\ell_x}(x)g_{\tilde{\lambda}\nu N n_y} 
T^{\ell_x}_{N N+1}R_{n_y\ell_y}(y)+ 
\sum_{n_x\in P} R_{N+1\ell_y}(y)g_{\tilde{\lambda}\nu n_x N}
T^{\ell_y}_{N N+1}R_{n_x\ell_x}(x) 
\right] 
({\bold N}^{-\frac 12})^{(11)}_{\tilde{\lambda}\lambda'}
\nonumber\\   
&\qquad+\sum_{\tilde{\nu}}\sum_{\tilde n_x \tilde n_y n_y\in P} 
R_{N+1\ell_x}(x)R_{n_y\ell_y}(y)T^{\ell_x}_{N+1 N}
\left(\frac{1}{2}\Lambda^{-\frac 12}_{\nu N n_y,\tilde{\nu}\tilde
    n_x\tilde n_y } 
+\frac{1}{2}\Lambda^{\frac 12}_{\nu N n_y,\tilde{\nu}\tilde n_x\tilde n_y} 
\right)
({\bold N}^{-\frac 12})^{(21)}_{\lambda'\tilde{\nu}\tilde n_x\tilde n_y}
\nonumber\\ 
&\qquad+\sum_{\tilde{\nu}}\sum_{\tilde n_x\tilde n_y n_x\in P} 
R_{n_x\ell_x}(x)R_{N+1\ell_y}(y)T^{\ell_y}_{N+1 N}
\left(\frac{1}{2}\Lambda^{-\frac 12}_{\nu n_x N,\tilde{\nu}\tilde
    n_x\tilde n_y} 
+\frac{1}{2}\Lambda^{\frac 12}_{\nu n_x N,\tilde{\nu}\tilde n_x \tilde n_y} 
\right)
({\bold N}^{-\frac 12})^{(21)}_{\lambda'\tilde{\nu}\tilde n_x\tilde n_y}\,, 
%\nonumber\\
\end{align} 
and 
\begin{align}
&\overline{\bold H}^{(22)}_{\nu  \nu^\prime}(x, y, x^\prime, y^\prime)   
=\frac{\delta(y-y')}{yy'}\delta_{\nu\nu'}T_{\nu}(x)\frac{\delta(x-x')}{xx'}
+\frac{\delta(x-x')}{xx'}\delta_{\nu\nu'}T_{\nu}(y)\frac{\delta(y-y')}{yy'}
\nonumber\\
&\qquad-\delta_{\nu\nu'}
\left[ \sum_{n_y\in P}\left( R_{N+1\ell_x}(x)T^{\ell_x}_{N+1N}R_{N\ell_x}(x')
+ R_{N\ell_x}(x)T^{\ell_x}_{NN+1}R_{N+1\ell_x}(x')\right) 
R_{n_y\ell_y}(y)R_{n_y\ell_y}(y')
\right.
\nonumber\\
&\qquad+\sum_{n_x\in P} R_{n_x\ell_x}(x)R_{n_x\ell_x}(x')\left(R_{N+1\ell_y}(y)T^{\ell_y}_{N+1N}R_{N\ell_y}(y') 
+R_{N\ell_y}(y)T^{\ell_y}_{NN+1}R_{N+1\ell_y}(y')\right) 
\nonumber\\
&\qquad+
\sum_{n_xn_yn_x'\in P}R_{n_x\ell_x}(x)T^{\ell_x}_{n_xn_x'}R_{n_x'\ell_x'}(x')R_{n_y\ell_y}(y)R_{n_y\ell_y'}(y')
\nonumber\\
&\qquad+\left.
\sum_{n_xn_yn_y'\in P}R_{n_x\ell_x}(x')R_{n_x\ell_x'}(x')
R_{n_y\ell_y}(y)T^{\ell_y}_{n_yn_y'}R_{n_y'\ell_y'}(y')
\right]\nonumber\\
&\qquad+\sum_{n_xn_yn_x'n_y'}
R_{n_x'\ell_x'}(x')R_{n_y'\ell_y'}(y')
R_{n_x\ell_x}(x)R_{n_y\ell_y}(y)
\overline{\bold H}^{(22)}_{\nu n_xn_y\nu'n_x'n_y'}
\nonumber\\
&\qquad+\sum_{\tilde{\lambda}}\left[\sum_{n_y\in P} R_{N+1\ell_x}(x)g_{\tilde{\lambda}\nu N n_y} 
T^{\ell_x}_{N N+1}R_{n_y\ell_y}(y)+ 
\sum_{n_x\in P} R_{N+1\ell_y}(y)g_{\tilde{\lambda}\nu n_x N}
T^{\ell_y}_{N N+1}R_{n_x\ell_x}(x) 
\right]\nonumber\\ 
&\qquad\times \sum_{n_x'n_y'\in
  P}R_{n_x'\ell_x'}(x')R_{n_y'\ell_y'}(y')({\bold N}^{-\frac 12})^{(12)}_{\tilde{\lambda}\nu'n_x'n_y'}
\nonumber\\    
&\qquad+\sum_{\tilde{\nu}}\sum_{\substack{\tilde n_x\tilde n_y\\n_yn_x'\\n_y'\in P}}  
R_{N+1\ell_x}(x)R_{n_y\ell_y}(y)T^{\ell_x}_{N+1 N}
\left(\frac{1}{2}\Lambda^{-\frac 12}_{\nu N n_y,\tilde{\nu}\tilde n_x \tilde n_y } 
+\frac{1}{2}\Lambda^{\frac 12}_{\nu N n_y,\tilde{\nu}\tilde n_x \tilde n_y } 
\right)
({\bold N}^{-\frac 12})^{(22)}_{\tilde{\nu}\tilde n_x\tilde n_y,\nu'n_x'n_y'} R_{n_x'\ell_x'}(x')R_{n_y'\ell_y'}(y')
\nonumber\\ 
&\qquad+\sum_{\tilde{\nu}}\sum_{\substack{\tilde n_x \tilde n_y \\n_x'n_y'\\n_x\in P}} 
R_{n_x\ell_x}(x)R_{N+1\ell_y}(y)T^{\ell_y}_{N+1 N}
\left(\frac{1}{2}\Lambda^{-\frac 12}_{\nu n_x N,\tilde{\nu}\tilde n_x \tilde n_y } 
+\frac{1}{2}\Lambda^{\frac 12}_{\nu n_x N,\tilde{\nu}\tilde n_x \tilde n_y } 
\right)
({\bold N}^{-\frac 12})^{(22)}_{\tilde{\nu}\tilde n_x\tilde n_y\nu'n_x'n_y'} R_{n_x'\ell_x'}(x')R_{n_y'\ell_y'}(y')
\nonumber\\
&\qquad+\sum_{\tilde{\lambda}}\sum_{n_xn_y}({\bold N}^{-\frac 12})^{(21)}_{\tilde{\lambda}\nu n_xn_y}
 R_{n_x\ell_x}(x)R_{n_y\ell_y}(y)
\nonumber\\
&\qquad\times
\left[\sum_{n_y'\in P} R_{N+1\ell_x'}(x')g_{\tilde{\lambda}\nu'N n_y'} 
T^{\ell_x'}_{N N+1}R_{n_y'\ell_y'}(y')+ 
\sum_{n_x'\in P} R_{N+1\ell_y'}(y')g_{\tilde{\lambda}\nu'n_x'N}
T^{\ell_y'}_{N N+1}R_{n_x'\ell_x'}(x') \right] \nonumber\\    
&\qquad+\sum_{\tilde{\nu}}\sum_{\substack{\tilde n_x \tilde n_y \\n_xn_y\\n_y'\in P}}R_{n_x\ell_x}(x)R_{n_y\ell_y}(y) 
({\bold N}^{-\frac 12})^{(22)}_{\nu n_x n_y, \tilde{\nu}\tilde n_x\tilde  n_y} 
\left(\frac{1}{2}\Lambda^{-\frac 12}_{\tilde{\nu}\tilde n_x \tilde n_y, \nu'N n_y'} 
+\frac{1}{2}\Lambda^{\frac 12}_{\tilde{\nu}\tilde n_x \tilde n_y, \nu'N n_y'} 
\right)
T^{\ell_x'}_{N N+1}R_{N+1\ell_x'}(x')R_{n_y'\ell_y'}(y')\nonumber\\ 
&\qquad+\sum_{\tilde{\nu}}\sum_{\substack{\tilde n_x \tilde n_y \\n_xn_y\\n_x'\in P}} R_{n_x\ell_x}(x)R_{n_y\ell_y}(y)
({\bold N}^{-\frac 12})^{(22)}_{\nu n_yn_y, \tilde{\nu}\tilde n_x\tilde n_y}
\left(\frac{1}{2}\Lambda^{-\frac 12}_{\tilde{\nu}\tilde n_x \tilde n_y, \nu'n_x'N} 
+\frac{1}{2}\Lambda^{\frac 12}_{\tilde{\nu}\tilde n_x \tilde n_y, \nu'n_x'N} 
\right)
T^{\ell_y'}_{N N+1}R_{N+1\ell_y'}(y')R_{n_x'\ell_x'}(x') \,, 
\end{align}   
where the subindex $N$ is the maximum size of the HO model space, which has been referred to as $N_{\rm max}$ throughout the paper,
$T^{\ell}_{nn'}=\langle n\ell|\hat T_{rel}|n'\ell\rangle$ are matrix
elements of the relative kinetic energy operator,  
and  $\Lambda$ represents the model-space norm kernel 
within the more limited formalism for the description of three-cluster dynamics
based solely on expansions over three-cluster channels states of the
type of Eq.~\eqref{eq:3bchannel} (see Eqs. (A3) - (A6) of Ref.~\cite{Quaglioni2013}). 

\section{Wave functions}
\label{ap:wf}
As described in section~\ref{sec:solutions}, instead of solving directly 
Eq.~\eqref{eq:NCSMC-eq}
we solve the set of orthogonalized Schr\"odinger equations Eq.~\eqref{eq:orthoH}.
Therefore, we obtain the orthogonalized vector of the expansion
coefficients $\overline{\bold C}^{\lambda}_{\nu x y}$ instead of the original
${\bold C}^{\lambda}_{\nu x y}$. These two arrays are related through 
Eq.~\eqref{eq:orthowf}, which can be inverted into

\begin{align}
        \label{eq:orthowf_inv}
        {\bold C}^{\lambda}_{\nu x y} & = \left[{\bold N}^{-\frac12} 
\overline{\bold C}\right]^{\lambda}_{\nu x y}%\\[2mm]
                                                  = \left(
                                                        \begin{array}{c}
                                                                {c}_{\lambda} \\
                                                                \chi_\nu(x,y)
                                                        \end{array}
                                                 \right)\,.
\end{align}

Therefore, we can recover the original ${\bold C}^{\lambda}_{\nu x y}$ through the following
expressions:
\begin{eqnarray}
{c}_{\lambda}&=&\sum_{\lambda'}({\bold N}^{-\frac12})^{(11)}_{\lambda\lambda'}\overline c_{\lambda'}  
+ \sum_{\nu}\iint dx x^2 dy y^2 ({\bold N}^{-\frac12})^{(12)}_{\lambda\nu x y} \overline\chi_{\nu}( x y) \nonumber\\
\chi_{\nu} (x y)&=& \sum_{\lambda} ({\bold N}^{-\frac12})^{(21)}_{\lambda\nu x y}\overline c_{\lambda}  
%\nonumber \\ 
+ \sum_{\nu'}\iint dx' x'^2 dy' y'^2 ({\bold N}^{-\frac12})^{(22)}_{\nu x y\nu' x' y'}
\overline\chi_{\nu'} (x' y')\,.
\end{eqnarray}
\section{Radii expressions}
\label{ap:radii}

The expectation value for the radii operators within the NCSMC wave function can be expressed in terms
of the cluster and composite bases as
%\begin{widetext}
\begin{align}
\label{radii}
\langle \Psi^{J^\pi T}| \hat r^2 |\Psi^{J^\pi T}\rangle 
&= \sum_{\lambda\lambda'} c_{\lambda}c_{\lambda'}\langle A\lambda J^{\pi}T|\hat r^2|A\lambda' J^{\pi}T\rangle 
\nonumber \\ 
&+\sum_{\lambda\nu'}c_{\lambda}\int dx'dy'x'^2y'^2 G_{\nu'}^{J^{\pi}T}(x',y')
\langle\phi_{\nu'x'y'}^{J^{\pi}T}
|\hat A_{\nu'}\hat r^2| A\lambda J^{\pi}T\rangle
\nonumber \\ 
&+\sum_{\lambda'\nu}c_{\lambda'}\int dxdyx^2y^2G_{\nu}^{J^{\pi}T}(x,y)\langle A\lambda' J^{\pi}T
|\hat r^2\hat A_{\nu}| \phi_{\nu xy}^{J^{\pi}T}\rangle
\nonumber \\  &
+\sum_{\nu\nu'}\iint dxdydx'dy'x^2y^2x'^2y'^2 G_{\nu}^{J^{\pi}T}(x,y)G_{\nu'}^{J^{\pi}T}(x',y')
\langle\phi_{\nu'x'y'}^{J^{\pi}T}|\hat A_{\nu'}\hat r^2\hat A_{\nu}|  \phi_{\nu xy}^{J^{\pi}T}\rangle, 
\end{align}
where, $\hat r^2$ represents either the matter or point proton radii operators. The root mean square radii are given  
by the square root of these matrix elements. 
Note that in Eq.~\eqref{radii} the first term corresponds to the expectation value within a NCSM calculation weighted by
the product of the discrete expansion amplitudes $c_{\lambda}$ and $c_{\lambda^\prime}$. This first term is calculated using the general expressions of the corresponding operators, 
however, the rest of the terms are calculated using the expressions that were derived in Sec.~\ref{sec:radius} considering the clusterization 
of the system, i.e., Eq.~\eqref{rad_3B} and the right side of Eq.~\eqref{eq:rpp} for the matter and point-proton radii, respectively.   
For the coupling terms, i.e., the second and third terms in Eq.~\eqref{radii},  mixed matrix   
elements are needed. We calculate these matrix elements by expanding, in an approximate way, the NCSM state into the cluster 
 basis. While this is in principle a rough approximation we can conclude {\it a posteriori} that the results are not  
significantly  affected by this approximation given that the contribution of these coupling terms in this first order is already 
very small compared to the other terms.  

When calculating the matter radius, Eq.~\eqref{radii} reduces to

\begin{align}
&\langle \Psi^{J^\pi T}|  r_m^2 |\Psi^{J^\pi T}\rangle=\nonumber\\  
&\quad= \sum_{\lambda\lambda'} c^{J^{\pi}T}_{ \lambda}   c^{J^{\pi}T}_{ \lambda'}
 \langle A \lambda J^{\pi}T| r_m^2| A \lambda' J^{\pi}T\rangle 
\nonumber\\
&\quad+\left(\frac{A-2}{A}\right)\sum_{\nu\nu^\prime} \langle A-a_{23}\, \alpha_1 I_1^{\pi_1}T_1|r_m^{2,{\rm core}}|A-a_{23}\, \alpha_1 I_1^{\pi_1}T_1\rangle     
 \iint dx dy x^2 y^2 W^{J^{\pi}T}_{\nu\nu'}(x,y) \nonumber\\  
&\quad+ 
\frac{1}{A}\sum_{\nu\nu'}
  \iint dx dy x^2 y^2 \rho^{2} W^{J^{\pi}T}_{\nu\nu'}(x,y)\,.
\end{align}

For the point-proton radius, Eq.~\eqref{eq:rpp} is valid given that the $core$ is the only charged 
cluster and has isospin zero. The expectation value is given by  

\begin{align}
&\langle \Psi^{J^\pi T}|  r_{pp}^2 |\Psi^{J^\pi T}\rangle=\nonumber\\  
&\qquad= \sum_{\lambda} c^{J^{\pi}T}_{ \lambda}   c^{J^{\pi}T}_{ \lambda'}  
 \langle A \lambda J^{\pi}T| r_{pp}^2| A \lambda' J^{\pi}T\rangle  \nonumber
\\
&\qquad+ \langle A-a_{23}\, \alpha_1 I_1^{\pi_1}T_1|r_{pp}^{2,{\rm core}}|A-a_{23}\, \alpha_1 I_1^{\pi_1}T_1\rangle     
 \iint dx dy x^2 y^2 W^{J^{\pi}T}_{\nu\nu'}(x,y) \nonumber\\ 
&\qquad+ 
\sqrt{\frac{2}{A(A-2)}}\sum_{\nu\nu'}
  \iint dx dy x^2 y^4  W^{J^{\pi}T}_{\nu\nu'}(x,y)\,. 
\end{align}
Here and in the equation above we have defined
\begin{align}
\nonumber
W^{J^{\pi}T}_{\nu\nu'}(x,y)&=\frac{1}{2}G^{J^{\pi}T}_{\nu' -}(x,y) G^{J^{\pi}T}_{\nu +}(x,y) %\\ \nonumber 
+\frac{1}{2}G^{J^{\pi}T}_{\nu'+}(x,y)G^{J^{\pi}T}_{\nu -}(x,y)\\ %\nonumber
&+\sum_{\lambda'} c_{\lambda'}^{J^{\pi}T}g^{J^{\pi}T}_{\lambda'\nu'}(x,y)G^{J^{\pi}T}_{\nu -}(x,y)%\\  
+G^{J^{\pi}T}_{\nu' -}(x,y)\sum_\lambda c_\lambda^{J^{\pi}T}g^{J^{\pi}T}_{\lambda\nu}(x,y), 
\end{align}
with
\begin{equation}
G^{J^{\pi}T}_{\nu \pm}(x,y)=\sum_{\nu'} \iint dx' dy' x'^2 y'^2 [\mathcal{N}^{J^{\pi}T}_{\nu\nu'}(x,y,x',y')]^{\pm1/2} \chi^{J^{\pi}T}_{\nu'}(x',y'). 
\end{equation}

\section{Parameters of the calculations}
\label{ap:parameters}
For completeness, in Tables~\ref{tab-lamda1.5} and ~\ref{tab-lamda2.0}  we list all parameters besides the HO model space size ($N_{\rm max}=12$) used for our best calculations 
for each $J^\pi T$ channel.
\begin{table}[h]
\caption{Parameters used for the calculations with $\lambda_{\rm{SRG}}$=1.5 fm$^{-1}$}
\begin{ruledtabular}  
\begin{tabular}{ c c c c c c} 
$J^\pi$ &  $N_{\mbox{ext}}$ &$K_{\mbox{max}}$ & $a$ (fm) & $n_s$ & $n_\alpha$   \\
\hline
0$^+$   & 	200	      &  40             & 45       &  125  & 40 \\
0$^-$   & 	70	      &  18             & 30       &  60   & 20 \\
1$^+$   &       70            &  19             & 30       &  60   & 30 \\
1$^-$   &       110           &  23             & 40       &  80   & 40\\
2$^+$   &       90            &  20             & 30       &  60   & 40 \\
2$^-$   &       70            &  18             & 30       &  60   & 20 \\
\end{tabular}
\end{ruledtabular}
\label{tab-lamda1.5} 
\end{table}

\begin{table}[h]
\caption{Parameters used for the calculations with $\lambda_{\rm{SRG}}$=2.0 fm$^{-1}$}
\begin{ruledtabular}  
\begin{tabular}{ c c c c c c} 
$J^\pi$ &  $N_{\mbox{ext}}$ &$K_{\mbox{max}}$ & $a$ (fm) & $n_s$ & $n_\alpha$   \\
\hline
0$^+$   & 	200	      &  40             & 45       &  150  & 50 \\
1$^+$   &       110           &  23             & 40       &  95   & 45 \\
1$^-$   &       110           &  23             & 40       &  95   & 45\\
2$^+$   &       90            &  20             & 30       &  60   & 40 \\
\end{tabular}
\end{ruledtabular} 
\label{tab-lamda2.0} 
\end{table}

\end{widetext}
\begin{acknowledgments}
Computing support for this work came from the Lawrence Livermore National Laboratory (LLNL) institutional Computing Grand Challenge program and from an INCITE Award on the Titan supercomputer of the Oak Ridge Leadership Computing Facility (OLCF) at ORNL. 
This article was prepared by LLNL under Contract DE-AC52-07NA27344. This material is based upon work supported by the U.S. Department of Energy, Office of Science, Office of Nuclear Physics, under Work Proposals No. SCW1158 and SCW0498, and by the Natural Sciences and Engineering Research Council of Canada (NSERC) Grants No. 401945-2011  and SAPIN-2016-00033. TRIUMF receives funding via a contribution through the 
Canadian National Research Council of Canada. 
\end{acknowledgments}

\bibliographystyle{apsrev4-1}
%\bibliography{biblio}

\end{document}